\documentclass[%
reprint,
superscriptaddress,
amsmath,
amssymb,
aps,
prb,
floatfix,
longbibliography
]{revtex4-1}
\usepackage{amssymb}
\usepackage{graphicx}
\usepackage{appendix}
\usepackage{color}
\usepackage{subfigure}
\usepackage[colorlinks=true,linkcolor=blue,filecolor=blue, urlcolor=blue,citecolor=blue]{hyperref}

\begin{document}

\preprint{APS/123-QED}

\graphicspath{{mainfigures/}}
\title{Site-selective insulating phase in twisted bilayer Hubbard model}
\author{Xiu-Cai Jiang}
\affiliation{Shanghai Key Laboratory of Special Artificial Microstructure Materials and Technology, School of Physics Science and engineering, Tongji University, Shanghai 200092, P.R. China}
\author{Ze Ruan}
\affiliation{Shanghai Key Laboratory of Special Artificial Microstructure Materials and Technology, School of Physics Science and engineering, Tongji University, Shanghai 200092, P.R. China}
\author{Yu-Zhong Zhang}
\email[Corresponding author.]{Email: yzzhang@tongji.edu.cn}
\affiliation{Shanghai Key Laboratory of Special Artificial Microstructure Materials and Technology, School of Physics Science and engineering, Tongji University, Shanghai 200092, P.R. China}

\date{\today}

\begin{abstract}
The paramagnetic phase diagrams of the half-filled Hubbard model on a twisted bilayer square lattice are investigated using coherent potential approximation. Besides the conventional metallic, band insulating, and Mott insulating phases, we find two site-selective insulating phases where certain sites exhibit band insulating behaviors while the others display Mott insulating behaviors. These phases are identified by the band gap, the double occupancy, the density of states, as well as the imaginary part of self-energy. Furthermore, we examine the effect of on-site potential on the stability of the site-selective insulating phases. Our results indicate that fruitful site-selective  phases can be engineered by twisting.
\end{abstract}

\maketitle
\section{Introduction}

Layered systems with twists have recently attracted intense attention due to the discoveries of numerous fascinating quantum phases, such as the Mott insulator\cite{cao2018correlated,chen2019evidence,regan2020mott,tang2020simulation,li2021continuous}, superconductivity\cite{cao2018unconventional,chen2019signatures}, and topological phases\cite{park2019higher,can2021high,eugenio2023twisted}. Meanwhile, the Hubbard model, employed to study various intriguing phases including the Mott insulator\cite{hubbard1964electron3}, orbital-selective phase\cite{koga2004orbital,de2009orbital,werner2007high,song2015possible}, bond-ordered insulator\cite{zhang2004dimerization,kancharla2007correlated}, superconductivity\cite{schulz1987superconductivity,maier2005systematic},  antiferromagnetism\cite{ogawa1975gutzwiller,schulz1990incommensurate,hirsch1989antiferromagnetism}, etc., has received significant interest for decades. Therefore, introducing twists in the Hubbard model may induce novel phases and are currently hot topics. Until now, much effort has been spent on the Hubbard model describing twisted transition metal dichalcogenides\cite{wu2018hubbard,pan2020band,pan2020quantum,wietek2022tunable,morales2022nonlocal} or twisted bilayer graphene\cite{huang2019antiferromagnetically}, predicting numbers of correlated phases. Besides, a few works investigate the Hubbard model on twisted bilayer square lattices but focus primarily on superconducting phase transitions\cite{lu2022doping,belanger2023doping}. However, the phase transitions among Mott insulator, band insulator, and metal in the Hubbard model on a twisted bilayer square lattice remains unexplored.

As we konw, even for the untwisted bilayer Hubbard model, the phase transitions at half filling are fascinating and widely investigated. Such a model can not only describe high-temperature cuprate superconductors\cite{fournier2010loss} but also be experimentally realized through fermionic atoms trapped in an optical lattice\cite{gall2021competing}. Quantum Monte Carlo simulations have revealed that the interlayer hopping suppresses intralayer long-range magnetic order in such a model on the square lattice\cite{scalettar1994magnetic}. Applying dynamical mean-field theory to the bilayer Hubbard model on a Bethe lattice, a smooth crossover between band and Mott insulators is discovered within the paramagnetic phase diagram\cite{fuhrmann2006mott}.  Furthermore, the magnetic phase diagram of this model on a square lattice is explored by cluster dynamical mean-field theory\cite{kancharla2007band} and quantum Monte Carlo simulations\cite{bouadim2008magnetic,mou2022bilayer}, where strong on-site Coulomb interaction localizes electrons with weak interlayer hopping, favoring a Mott insulator, while strong interlayer hopping opens a gap between the bonding and antibonding bands, resulting in a band insulator. Although these works\cite{kancharla2007band,bouadim2008magnetic,mou2022bilayer} also suggest a paramagnetic metallic phase in the magnetic phase diagram when both on-site Coulomb interaction and interlayer hopping are weak, further investigations clarify it as an antiferromagnetic insulating phase due to the perfect nesting property of the Fermi surface within the noninteracting system\cite{golor2014ground,ruger2014phase}. In contrast, a metallic phase can appear in the paramagnetic phase diagram with weak on-site Coulomb interaction present\cite{ruger2014phase}. In addition to these findings, superconductivity\cite{lanata2009superconductivity,zhai2009antiferromagnetically,maier2011pair,matsumoto2020strongly,kato2020many,karakuzu2021superconductivity,iwano2022superconductivity}, non-Fermi liquid\cite{lee2014competition}, density-ordered phase\cite{vanhala2015superfluidity}, and superfluid phases\cite{vanhala2015superfluidity} have also been reported in the bilayer Hubbard model or its extended versions as doping away from half filling.

When the twist is applied to the layered systems, the well-known moir{\'e} pattern emerges, giving rise to distinct hoppings present at the inequivalent sublattices due to different atomic environments. In fact, distinct hoppings on different sublattices may cause a site-selective phase. A typical example is that bond-length disproportionation, corresponding to inequivalent sublattices having distinct hoppings, leads to a site-selective insulating phase in $R$NiO$_{3}$ ($R$=Sm, Eu, Y, or Lu), where certain Ni atoms exhibit a magnetic Mott insulating state while the remaining Ni atoms form a singlet insulating state\cite{park2012site}. Moreover, site-selective magnetic phases have also been reported in other materials with inequivalent sublattices\cite{shimizu2015site,yu2014site}. Given that the interlayer hoppings favor a band insulator in bilayer square lattice, the instabilities of the band insulating state against the Mott insulator on inequivalent sublattices of twisted bilayer square lattice may be different in the presence of a strong on-site Coulomb interaction owing to distinct interlayer hoppings. Therefore, a site-selective insulating phase may probably appear in such a twisted system, characterized by some sites entering a Mott insulating state while others remain a band insulating state.

In this paper, we aim to point out the presence of site-selective insulating phases in the half-filled Hubbard model on a twisted bilayer square lattice. To this end, taking the case with a twisted angle of $\theta=53.13^\circ$ as an example, we investigate the paramagnetic phase diagrams of this model under the combined effect of on-site Coulomb interaction and various interlayer hoppings using coherent potential approximation (CPA). Interestingly, we obtain two site-selective insulating phases, where certain sites exhibit band insulating states while the rests display Mott insulating states, when one type of the interlayer hoppings is strong while the remaining interlayer hoppings are weak if the strong on-site Coulomb interaction is involved. In addition, we also discover Mott insulating, band insulating, and metallic phases in this model, which have been observed in the untwisted bilayer Hubbard model as well. The Mott insulating phase
emerges when strong on-site Coulomb interaction wins over all the weak interlayer hoppings. In contrast, strong $t_{\bot}$, either in cooperation with strong $t_{1}$ or strong $t_{2}$, favors the band insulating phase [the definitions of $t_{\bot}$, $t_{1}$, and $t_{2}$ can be found in Fig.\ref{structure}]. Furthermore, we have illustrated that the site-selective insulating phases are stable even in the presence of a moderate site-dependent on-site potential. Our findings not only demonstrate a fascinating phenomenon that the band insulating state and Mott insulating state can coexist in twisted strongly correlated systems but also suggest twist as an effective approach to access a site-selective phase.

The rest of this paper is organized as follows: Sec.~\ref{Model-method} describes the details of the structure, the model, and the method we used. Sec.~\ref{Results} demonstrates our primary results, including the paramagnetic phase diagrams under different parameters, the gaps and double occupancies as functions of various interlayer hoppings, the density of states (DOS), the imaginary part of self-energy, the DOS at the Fermi level varied with the Lorentzian broadening factor, as well as the effect of an on-site potential difference on the stability of the site-selective phases. Sec.~\ref{DISCUSSION} includes a discussion of our results and Sec.~\ref{Conclusion} concludes with a summary.

\section{model and method}\label{Model-method}
\begin{figure}[htbp]
\includegraphics[width=0.26\textwidth]{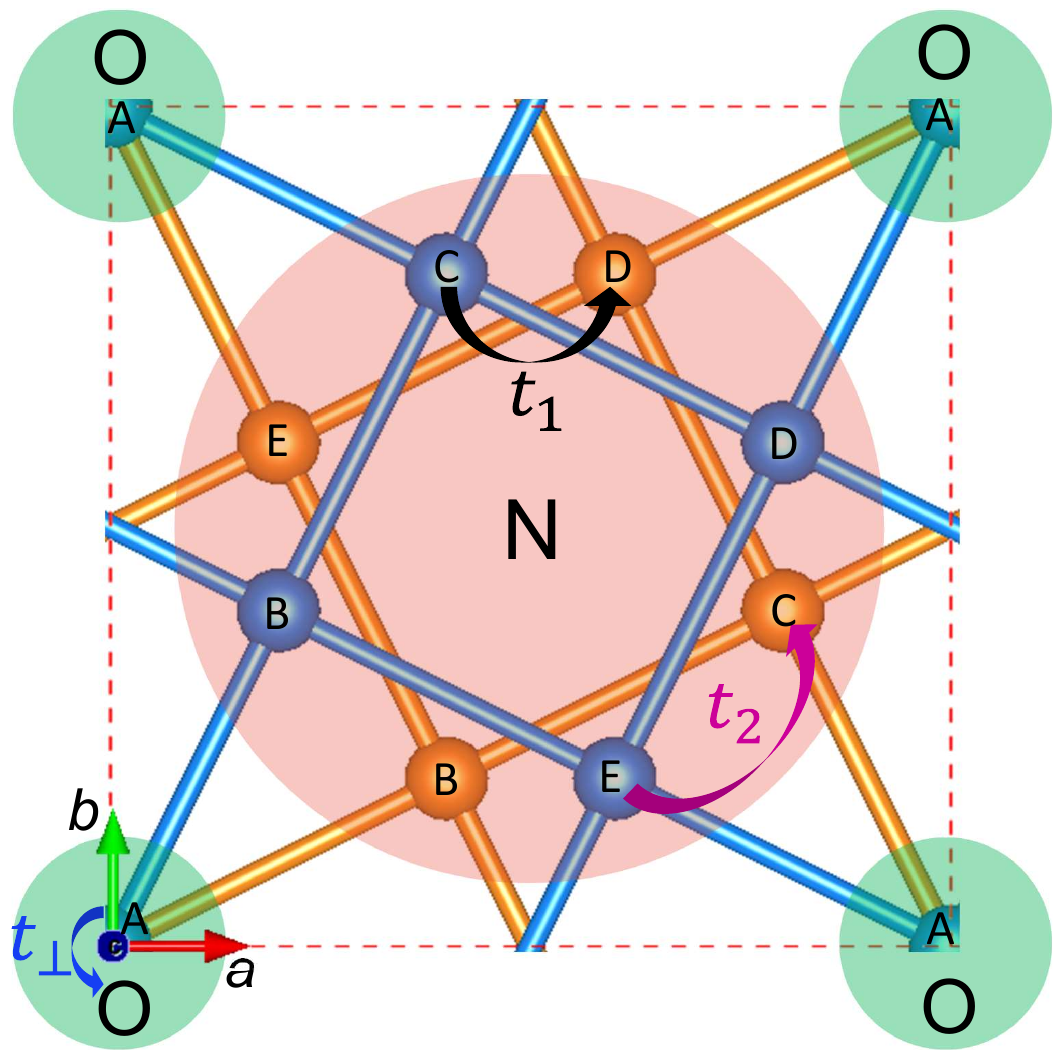}
\caption{ The structure of bilayer square lattice (top view) with a twisted angle of $\theta=53.13^\circ$, where different interlayer hoppings are also shown, including interlayer nearest-neighbor hopping $t_{\perp}$, interlayer next-nearest-neighbor hopping $t_{1}$, and interlayer third-nearest-neighbor hopping $t_{2}$. There are two types of inequivalent sites within the supercell, containing 2 overlapping sites and 8 non-overlapping sites (as viewed from the top), which are located in region O (green) and region N (light orange), respectively.}
\label{structure}
\end{figure}
To demonstrate the presence of site-selective insulating phases in the half-filled Hubbard model on a twisted bilayer square lattice, we take the case with a twisted angle of $\theta=53.13^\circ$ as an example since it is the smallest commensurate structure of twisted bilayer square lattice as presented in Fig.~\ref{structure} and has also been used to study the superconducting phase transitions\cite{lu2022doping}. According to the atomic environment, two types of inequivalent sites are distinguished within the supercell, containing 2 overlapping sites and 8 non-overlapping sites (as viewed from the top), which are located at region O (green) and region N (light orange), respectively. Then, the Hamiltonian can be written as
\begin{eqnarray}
H=H_{k}+H_{\bot}+H_{\Delta}+H_{\mu}+H_{U}\label{HubbardModel}
\end{eqnarray}
with
\begin{equation}
\begin{aligned}
H_{k}=&-t_{0}\sum\limits_{m\sigma}\sum\limits_{\langle{is,js^{\prime}}\rangle}C_{ism\sigma}^{\dag}C_{js^{\prime}m\sigma},\\
H_{\bot}=&-t_{\bot}\sum\limits_{i\sigma}(C_{iA1\sigma}^{\dag}C_{iA2\sigma}+H.c.)\\
&-t_{1}\sum\limits_{i\sigma}\sum\limits_{\langle\langle{s,s^{\prime}}\rangle\rangle}(C_{is1\sigma}^{\dag}C_{is^{\prime}2\sigma}+H.c.)\\
&-t_{2}\sum\limits_{i\sigma}\sum\limits_{\langle\langle\langle{s,s^{\prime}}\rangle\rangle\rangle}(C_{is1\sigma}^{\dag}C_{is^{\prime}2\sigma}+H.c.),\\
H_{\Delta}=&\Delta_O\sum\limits_{im\sigma}n_{iAm\sigma}+\Delta_N\sum\limits_{im\sigma}\sum\limits_{s\in{N}}n_{ism\sigma},\\
H_{\mu}=&-\mu\sum\limits_{ism\sigma}n_{ism\sigma},\\
H_{U}=&U\sum\limits_{ism}n_{ism\uparrow}n_{ism\downarrow},
\end{aligned}
\end{equation}
where $H_{k}$ is the Hamiltonian of the intralayer nearest-neighbor hopping. $H_{\bot}$ is the Hamiltonian describing interlayer nearest-neighbor, next-nearest-neighbor, and third-nearest-neighbor hoppings. $H_{\Delta}$ and $H_{\mu}$ denote separately the energies of on-site potential and chemical potential. $H_{U}$ depicts the on-site Coulomb repulsive interaction between spin-up and spin-down electrons.
Here, $i(j)$,  $s(s^{\prime})$, $m$, and $\sigma$ denote separately the cell, sublattice, layer, and spin indexes. $\langle{is,js^{\prime}}\rangle$, $\langle{\langle{s,s^{\prime}}\rangle}\rangle$, and $\langle\langle{\langle{s,s^{\prime}}\rangle}\rangle\rangle$ stand for the summations over the intralayer nearest-neighbor sites, interlayer next-nearest-neighbor sites, and interlayer third-nearest-neighbor sites, respectively. $t_{0}$ and $t_{\bot}$($t_{1}, t_{2}$) represent individually the intralayer nearest-neighbor hopping integral and interlayer nearest(next-nearest, third-nearest)-neighbor hopping integral. $\Delta_O$ and $\Delta_N$ are on-site potentials of the inequivalent sublattices. $\mu$ is the chemical potential, and $U$ is the on-site Coulomb repulsive interaction.

We now introduce how to employ the CPA to solve this many-body Hamiltonian. Hubbard views the electron correlation problem as a
disordered alloy where an electron with spin $\sigma$ moving in the system encounters either a potential of $U$ at a site with a spin $\bar{\sigma}$ present or $0$ without\cite{Hubbard1963Electron1}. Then, the alloy analogy of this Hubbard model has the following form
\begin{eqnarray}
H_{A}=H_{k}+H_{\bot}+H_{\Delta}+H_{\mu}+\sum\limits_{ism\sigma}E_{ism\sigma}n_{ism\sigma},\label{alloyModel}
\end{eqnarray}
where $E_{ism\sigma}$ is a disordered potential depending on the presence of a spin $\bar{\sigma}$. Specifically, $E_{ism\sigma}=U$ with a probability of $P_U=\langle{n_{ism\bar{\sigma}}}\rangle$ or $E_{ism\sigma}=0$ with a probability of $P_0=1-\langle{n_{ism\bar{\sigma}}}\rangle$. The Green's function of this disordered model necessitates the computation of an average overall possible disordered configurations. However, performing this calculation exactly is impossible, and the CPA should be employed to solve this alloy problem\cite{soven1967coherent,velicky1968single,elliott1974theory}. Within the framework of the CPA, the disordered alloy is self-consistently mapped into an effective medium, explicitly, the disordered potential $E_{ism\sigma}$ is substituted with an energy-dependent, site-diagonal, and translationally invariant self-energy $\Sigma_{sm\sigma}$. Then, the Hamiltonian of the effective medium within the CPA becomes
\begin{eqnarray}
H_{eff}=H_{k}+H_{\bot}+H_{\Delta}+H_{\mu}+\sum\limits_{ism\sigma}\Sigma_{sm\sigma}n_{ism\sigma}.\label{CPAModel}
\end{eqnarray}
The detailed mapping from model \eqref{alloyModel} to model \eqref{CPAModel} is given in Appendix \ref{mapping}. Noticeably, despite some inherent limitations\cite{gebhard1997metal}, the CPA remains valuable as a reliable and computationally cheap method for capturing the phase transitions among a band insulator, metal, and Mott insulator in many-body systems. For example, the CPA successfully reproduces the phase diagram of ionic Hubbard model at half filling\cite{hoang2010metal,rowlands2014inclusion}, the critical on-site Coulomb interaction for Mott transition on the honeycomb lattice at half filling obtained by the CPA\cite{rowlands2014disappearance} is consistent with the results of the quantum Monte Carlo simulations\cite{assaad2013pinning,sorella2012absence,toldin2015fermionic} and cluster dynamical mean field theory\cite{wu2010interacting,liebsch2011correlated},
the experimental discrepancies of the gap in both bilayer graphene\cite{mak2009observation,weitz2010broken} and graphene/h-BN heterostructure\cite{hunt2013massive,chen2014observation} have been successfully understood by employing the CPA to investigate their phase diagrams\cite{xu2016gate,xu2016interaction}.
\section{results}\label{Results}
\begin{figure}[htbp]
\includegraphics[width=0.48\textwidth,height=0.45\textwidth]{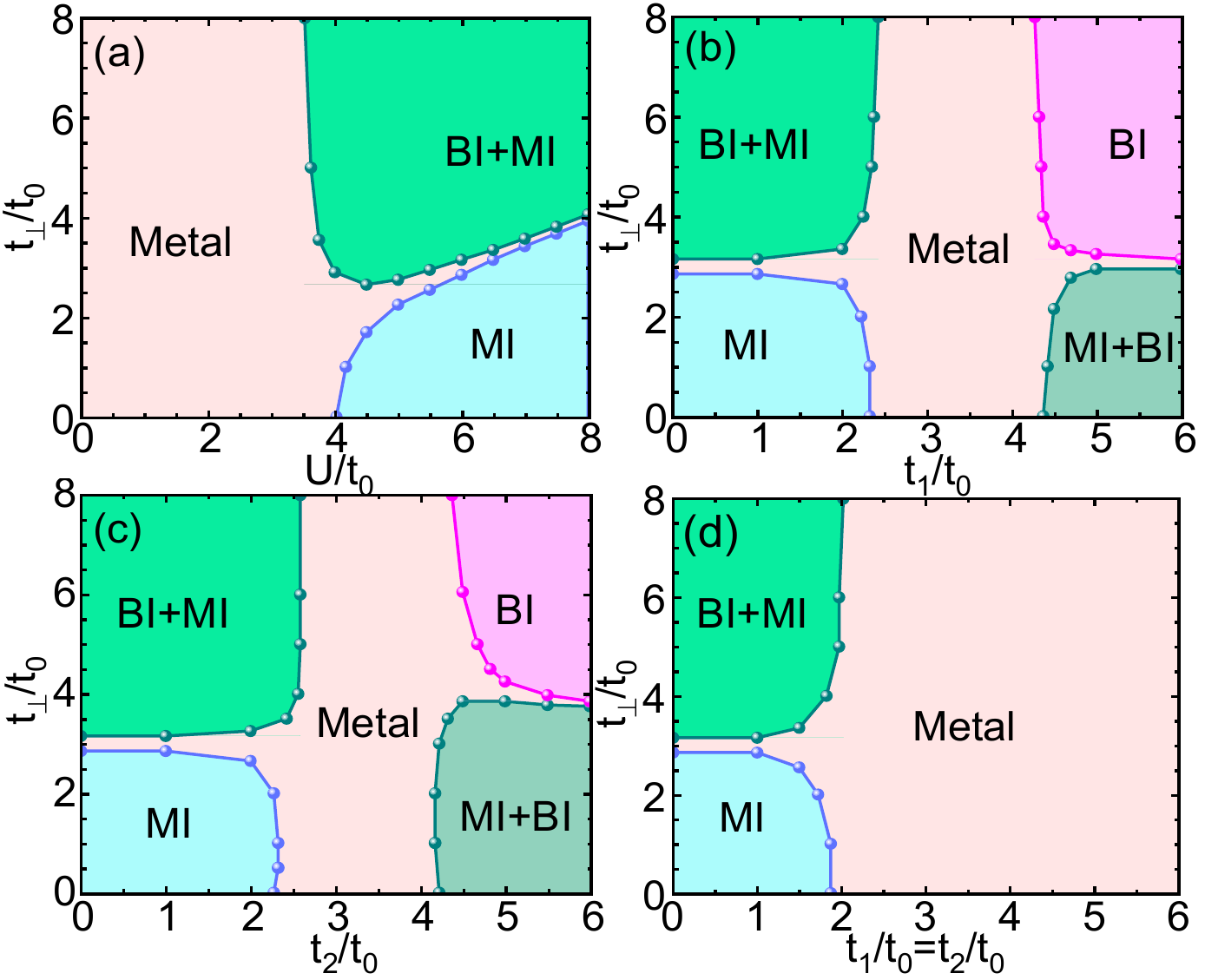}
\caption{The paramagnetic phase diagrams of the half-filled Hubbard model on twisted bilayer square lattice, involving (a) the $U/t_0-t_{\perp}/t_0$ plane at $t_1/t_0=t_2/t_0=0$, (b) the $t_1/t_0-t_{\perp}/t_0$ plane when $U=6t_0$ and $t_2/t_0=0$, (c) the $t_2/t_0-t_{\perp}/t_0$ plane when $U=6t_0$ and $t_1/t_0=0$, and (d) the $(t_1/t_0=t_2/t_0)-t_{\perp}/t_0$ plane at $U=6t_0$. $t_{0}$ and $t_{\bot}$ ($t_{1}, t_{2}$) are individually the intralayer and interlayer nearest (next-nearest, third-nearest)-neighbor hopping integrals. BI, MI, and BI+MI (MI+BI) denote the band insulating phase, Mott insulating phase, and site-selective insulating phase, respectively. Noticeably, BI+MI and MI+BI are two distinct site-selective insulating phases. Specifically, BI+MI (MI+BI) describes a site-selective insulating phase where the overlapping sites (located at region O) exhibit a band (Mott) insulating state while the non-overlapping sites (located at region N) manifest a Mott (band) insulating state.}
\label{phase-diagram-U}
\end{figure}

Now, we will demonstrate the presence of site-selective insulating phases in the paramagnetic phase diagrams of the half-filled Hubbard model on this twisted bilayer square lattice. To this end, we employ the CPA to calculate the paramagnetic phase diagrams of this model under the combined effect of on-site Coulomb interaction and various interlayer hoppings, where three interlayer hoppings are concerned, including interlayer nearest-neighbor hopping $t_{\perp}$, interlayer next-nearest-neighbor hopping $t_{1}$, and interlayer third-nearest-neighbor hopping $t_{2}$. Figure \ref{phase-diagram-U}(a) illustrates the phase diagram in the $U/t_0-t_{\perp}/t_0$ plane at $t_1/t_0=t_2/t_0=0$. As can be seen, when interlayer hoppings are absent, equivalent to two unrelated monolayer square lattices, the system undergoes a phase transition from metal to Mott insulator with increasing on-site Coulomb interaction, consistent with the results obtained by other methods\cite{groeber1998paramagnetic,gull2013superconductivity,ruger2014phase}. Thus, the CPA provides reliable results for two irrelevant monolayer square lattices, and we go on with the case of interlayer hoppings present. Remarkably, a site-selective insulating phase (BI+MI), where overlapping sites exhibit a band insulating state while non-overlapping sites display a Mott insulating state, emerges at the region of strong on-site Coulomb interaction if the interlayer nearest-neighbor hopping $t_{\perp}$ exceeds a critical value. This is because a strong $t_{\perp}$ generates interlayer singlets at the overlapping sites, corresponding to the appearance of a band insulating state there\cite{mou2022bilayer,ruger2014phase}, while a strong on-site Coulomb interaction $U$ stabilizes a Mott insulator state at non-overlapping sites due to the lack of interlayer hoppings.

Figure \ref{phase-diagram-U}(b) and \ref{phase-diagram-U}(c) demonstrate the phase diagrams in the $t_1/t_0-t_{\perp}/t_0$ plane at $t_2/t_0=0$ and $t_2/t_0-t_{\perp}/t_0$ plane at $t_1/t_0=0$, respectively, under a strong on-site Coulomb interaction of $U=6t_0$. Interestingly, two distinct site-selective insulating phases are observed in both phase diagrams, specifically BI+MI and MI+BI, where MI+BI is the counterpart phase of BI+MI. In MI+BI, a Mott (band) insulating state is replaced by a band (Mott) insulating state at specific sites compared with BI+MI.  Besides, despite slight differences in the phase boundaries, both phase diagrams contain the same phases. This happens because increasing either $t_1$ or $t_2$ will destroy the Mott insulating state at non-overlapping sites and subsequently form a band insulating state there. Consequently, as either $t_1$ or $t_2$ is increased, the system experiences phase transitions from the Mott insulating phase to a metallic phase and then into MI+BI for a weak $t_{\perp}$ while it evolves from BI+MI to a metallic phase and then into the band insulating phase for a strong $t_{\perp}$.

The phase diagram in the $(t_1/t_0=t_2/t_0)-t_{\perp}/t_0$ plane are also investigated when $U=6t_0$ as illustrated in Fig.\ref{phase-diagram-U}(d). Apparently, the enhancement of both $t_{1}$ and $t_{2}$ also destroys the Mott insulating state at non-overlapping sites. As a result, although the critical values of $t_{1}$ and $t_{2}$ resulting in the phase transition decrease, the phase transitions in Fig.\ref{phase-diagram-U}(d) are comparable to those in Fig.\ref{phase-diagram-U}(b) and \ref{phase-diagram-U}(c) when both $t_{1}$ and $t_{2}$ are weak. However, in the region where both $t_{1}$ and $t_{2}$ are strong, neither BI nor MI+BI will occur as the interlayer singlets between non-overlapping sites fail to form when strong $t_{1}$ and strong $t_{2}$ present simultaneously.

In brief, in this twisted bilayer Hubbard model, we not only observe the Mott insulating, band insulating, and metallic phases proposed in the untwisted case but also identify two site-selective phases with the coexistence of band and Mott insulating states. These findings suggest twist as an effective approach to access a site-selective phase in strongly correlated systems.

\begin{figure}[htp]
\includegraphics[width=0.44\textwidth,height=0.35\textwidth]{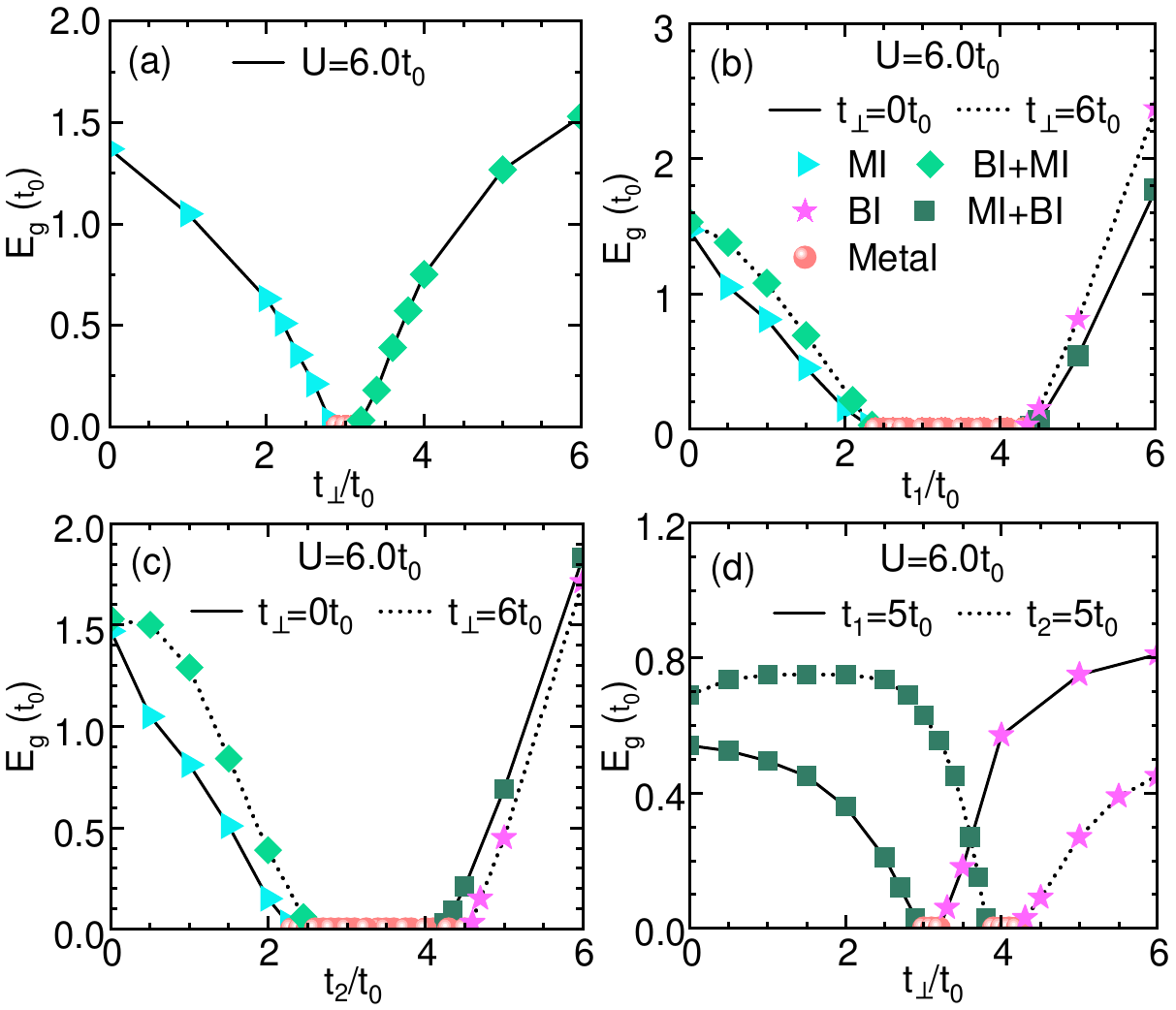}
\caption{The band gaps as functions of various interlayer hoppings. (a) The band gap varies with interlayer nearest-neighbor hopping $t_{\perp}$ when $t_1/t_0=t_2/t_0=0$ and $U=6t_0$. (b) Band gaps as functions of interlayer next-nearest-neighbor hopping $t_{1}$ for two indicated $t_{\perp}$ values, namely, $0t_0$ (solid line) and $6t_0$ (dotted line), where $t_2/t_0=0$ and $U=6t_0$ are used. (c) Band gaps as functions of interlayer third-nearest-neighbor hopping $t_{2}$ for two specified $t_{\perp}$ values, $0t_0$ (solid line) and $6t_0$ (dotted line), where $t_1/t_0=0$ and $U=6t_0$ are adopted. (d) The evolution of band gaps with interlayer nearest-neighbor hopping $t_{\perp}$ when $t_{1}=5t_0$ and $U=6t_0$ (solid line) as well as when $t_{2}=5t_0$ and $U=6t_0$ (dotted line). The green diamonds and malachite green squares represent two distinct site-selective phases. The pink pentagram, cyan triangle, and orange ball denote the band insulating, Mott insulating, and metallic phases, respectively.}
\label{gaps-tps}
\end{figure}
\begin{figure}[htp]
\includegraphics[width=0.44\textwidth,height=0.36\textwidth]{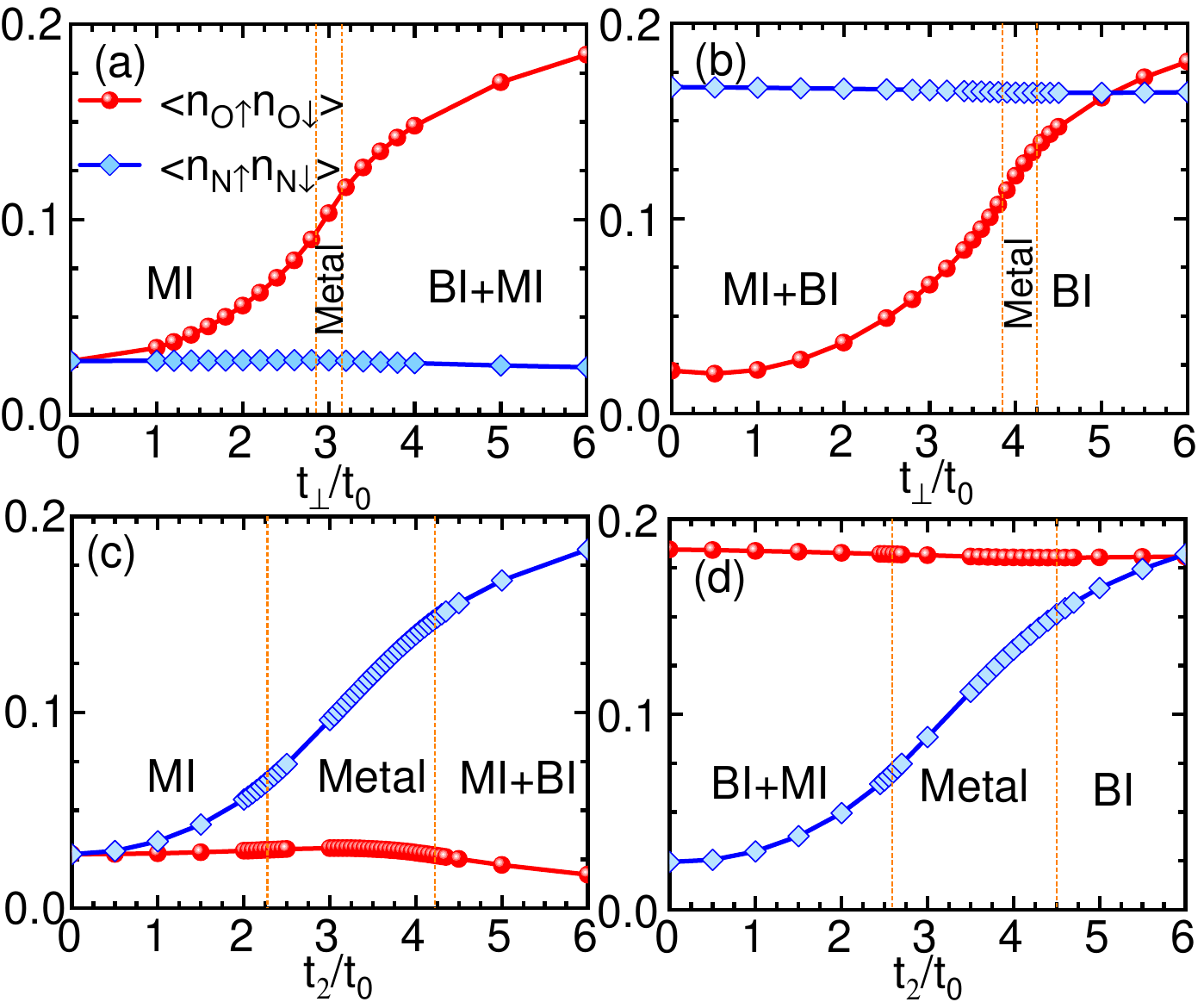}
\caption{The double occupancies on different sublattices as functions of various interlayer hoppings. (a) The double occupancies on different sublattices varies with interlayer nearest-neighbor hopping $t_\perp$ when $t_1/t_0=t_2/t_0=0$ and $U=6t_0$. (b) The double occupancies on different sublattices as functions of $t_\perp$ when $t_2/t_0=5$, $t_1/t_0=0$, and $U=6t_0$. (c) The double occupancies on different sublattices varies with interlayer next-nearest-neighbor hopping $t_2$ when $t_\perp/t_0=t_1/t_0=0$ and $U=6t_0$. (d) The double occupancies on different sublattices as functions of $t_2$ when $t_\perp/t_0=6$, $t_1/t_0=0$, and $U=6t_0$.}
\label{double-occu}
\end{figure}
\begin{figure*}[htbp]
\includegraphics[width=0.98\textwidth,height=0.54\textwidth]{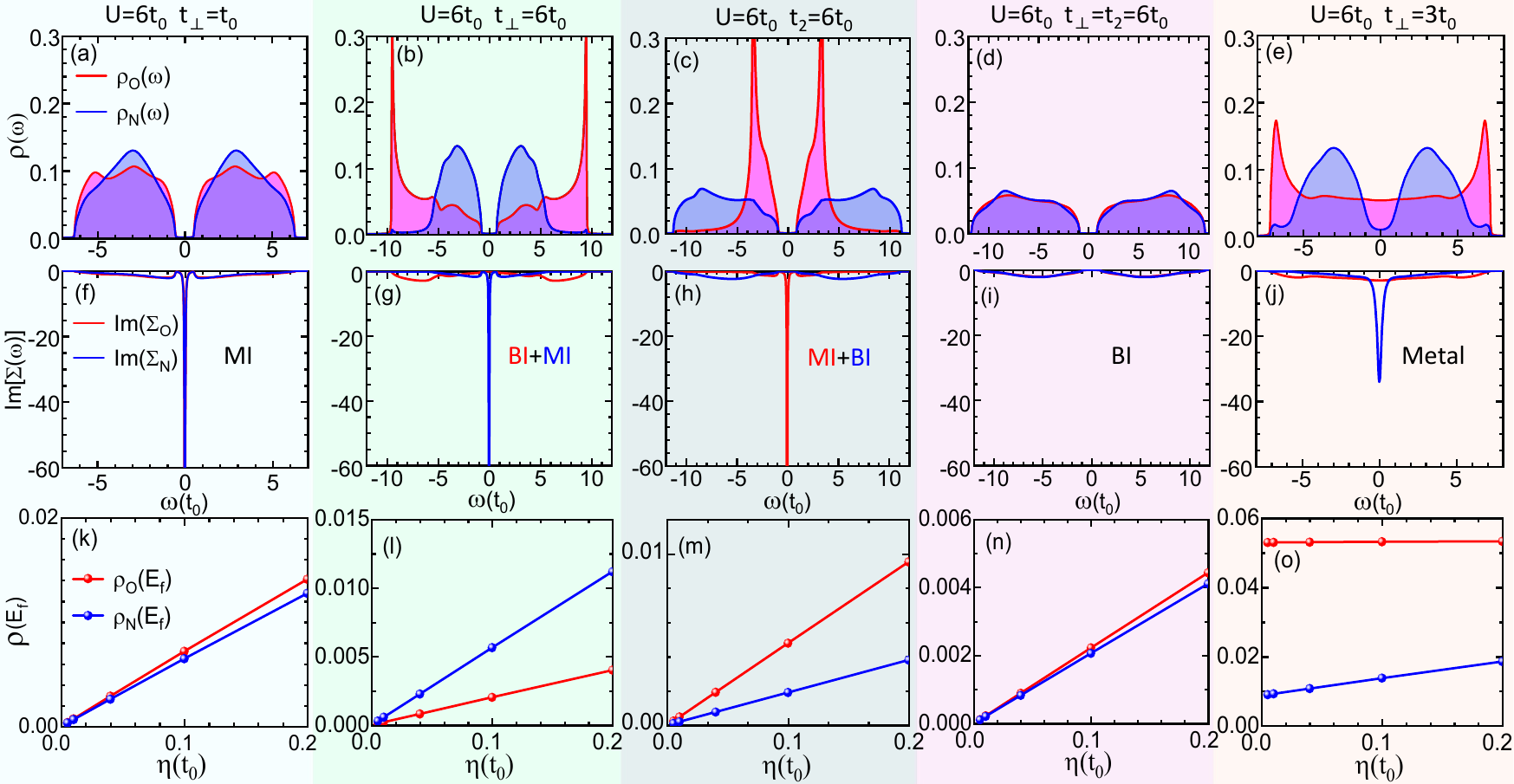}
\caption{The upper panel [(a)-(e)] depicts the DOS at various parameter points, where $\rho_O(\omega)$ (pink shadow) and $\rho_N(\omega)$ (blue shadow) represent the DOS at each site in regions O and N of the supercell, respectively. The middle panel [(f)-(j)] displays the imaginary part of self-energies under various parameter points, where $\rm{Im}(\Sigma_O)$ (red line) and $\rm{Im}(\Sigma_N)$ (blue line) describe the imaginary part of self-energy at each site in regions O and N, respectively. The lower panel [(k)-(o)] shows the DOS at Fermi level varied with the Lorentzian broadening factor $\eta$ under various parameter points, where $\rho_O(E_f)$ (red line) and $\rho_N(E_f)$ (blue line) separately denote the DOS at Fermi level of each site in regions O and N. Specifically, (a), (f), and (k) correspond to a Mott insulating phase, where $U=6t_0$, $t_{\perp}=t_0$, and $t_1/t_0=t_2/t_0=0$. (b), (g), (l) ($U=6t_0$, $t_{\perp}=6t_0$, and $t_1/t_0=t_2/t_0=0$) as well as (c), (h), (m) ($U=6t_0$, $t_{2}=6t_0$, and $t_1/t_0=t_\perp/t_0=0$) depict two distinct site-selective insulating phases. (d), (i), and (n) characterize a band insulating phase when $U=6t_0$, $t_2=t_{\perp}=6t_0$, and $t_1/t_0=0$. (e), (j), and (o) describe a metallic phase under the parameters of $U=6t_0$, $t_{\perp}=3t_0$, and $t_1/t_0=t_2/t_0=0$. A Lorentzian broadening factor of $\eta=0.01t_0$ is used in the upper and middle panels. As the MI+BI and BI in the $t_1/t_0-t_{\perp}/t_0$ plane are the same as those in the $t_2/t_0-t_{\perp}/t_0$ plane, we only present the corresponding physical quantities for the MI+BI and BI in the latter here.}
\label{DOS-Selfenergy-etas}
\end{figure*}

Next, we proceed to explain how the phases in the paramagnetic phase diagrams (Fig.\ref{phase-diagram-U}) are determined. As the trend in the band gap with interlayer hopping provides a valuable distinguishing characteristic for the phases in the untwisted case\cite{kancharla2007band}, we plot the evolution of the band gap in this twisted system under various interlayer hoppings in Fig.\ref{gaps-tps}, where a sufficient strong on-site Coulomb interaction with $U=6t_0$ is adopted. We discover that the band gap first closes and then reopens as $t_{\perp}$ increases in the absence of other interlayer hoppings [Fig.\ref{gaps-tps}(a)], confirming the presence of two insulating phases separated by a metallic phase within the $U/t_0-t_{\perp}/t_0$ plane [Fig.\ref{phase-diagram-U}(a)]. By inspecting the effect of $t_{1}$ [Fig.\ref{gaps-tps}(b)] and that of $t_{2}$ [Fig.\ref{gaps-tps}(c)] on the band gap of these two insulating phases, four insulating phases are distinguishable in both $t_1/t_0-t_{\perp}/t_0$ plane [Fig.\ref{phase-diagram-U}(b)] and $t_2/t_0-t_{\perp}/t_0$ plane [Fig.\ref{phase-diagram-U}(c)]. This can be understood by the following facts: the band gaps of two insulating phases observed in the $U/t_0-t_{\perp}/t_0$ plane gradually vanish as $t_{1}$ or $t_{2}$ is increased, indicating their disappearances, meanwhile, the gaps reopen and increase for strong values of both $t_{1}$ and $t_{2}$, implying the occurrence of additional insulating phases which are confirmed by Fig.\ref{gaps-tps}(d) as two new insulating phases.

Besides the behavior of the band gap, we further confirm the presence of four distinct insulating phases by analyzing the double occupancy. Fig.\ref{double-occu} demonstrates the double occupancies on inequivalent sites of four line slices within the phase diagram plane presented in Fig.\ref{phase-diagram-U}(c) which includes all the phases discovered. As can be seen, in the insulating phase located at the region where both $t_{\perp}$ and $t_{2}$ are weak, all double occupancies are suppressed to small values [MI region in Fig.\ref{double-occu}(a) and Fig.\ref{double-occu}(c)]. For this insulating phase, increasing either $t_{\perp}$ or $t_{2}$ will cause an increase in double occupancies on corresponding sites due to the formation of interlayer singlets there, whereas the double occupancies on the other sites remain nearly unchanged. This ultimately results in the emergence of two distinct insulating phases characterized by the coexistence of both small and large double occupancies [BI+MI region in Fig.\ref{double-occu}(a) and MI+BI region in Fig.\ref{double-occu}(c)]. Besides, all double occupancies exhibit relatively large values in the insulating phase where both $t_{\perp}$ and $t_{2}$ are strong [BI region in Fig.\ref{double-occu}(b) and Fig.\ref{double-occu}(d)]. Therefore, based on the behavior of the band gap and double occupancy, four distinct insulating phases are identified.

In order to clarify the natures of the aforementioned four insulating phases, we select one parameter point within each insulating phase [The parameter points are indicated above Fig.\ref{DOS-Selfenergy-etas}(a)-\ref{DOS-Selfenergy-etas}(d), where the unspecified parameters are all set to 0.] to calculate the corresponding DOS, imaginary part of self-energy, and Lorentzian broadening factor $\eta$ dependence of DOS at the Fermi level, as shown in Fig.\ref{DOS-Selfenergy-etas}, where the imaginary part of self-energy is used to identify the Mott and band insulating states while the DOS serves to distinguish the insulating phases from a metallic phase. Obviously, the opening of the band gap within the DOS in Figs. \ref{DOS-Selfenergy-etas}(a)-\ref{DOS-Selfenergy-etas}(d) and the disappearance of both $\rho_O(E_f)$ and $\rho_N(E_f)$ as $\eta$ approaches zero in Figs. \ref{DOS-Selfenergy-etas}(k)-\ref{DOS-Selfenergy-etas}(n) further confirm the insulating behaviors of these phases.

We now focus on the imaginary part of their self-energies. It is apparent from Fig.\ref{DOS-Selfenergy-etas}(f) that both $\rm{Im}(\Sigma_O)$ and $\rm{Im}(\Sigma_N)$ diverge proximity to zero frequency, which is a typical character of a Mott insulating phase\cite{song2015possible,song2017distinct}. In contrast, both $\rm{Im}(\Sigma_O)$ and $\rm{Im}(\Sigma_N)$ vanish at $\omega=0$ in Fig.\ref{DOS-Selfenergy-etas}(i), clearly indicating the occurrence of a band insulating phase. Surprisingly, $\rm{Im}(\Sigma_O)$ and $\rm{Im}(\Sigma_N)$ exhibit distinct behaviors at zero frequency for both Fig.\ref{DOS-Selfenergy-etas}(g) and Fig.\ref{DOS-Selfenergy-etas}(h), where $\rm{Im}(\Sigma_N)$ diverges but $\rm{Im}(\Sigma_O)$ vanishes in Fig.\ref{DOS-Selfenergy-etas}(g) while $\rm{Im}(\Sigma_N)$ vanishes but $\rm{Im}(\Sigma_O)$ diverges in Fig.\ref{DOS-Selfenergy-etas}(h), suggesting the presence of two distinct site-selective phases with the coexistence of band and Mott insulating states in the system.

For comparison, we also calculate the same physical quantities for a metallic phase. Apparently, the closure of the gap in the DOS [\ref{DOS-Selfenergy-etas}(e)] and the finite values of both $\rho_O(E_f)$ and $\rho_N(E_f)$ as $\eta$ approaches zero [\ref{DOS-Selfenergy-etas}(o)] are key features of a metallic phase. It is necessary to mention that the nonzero imaginary part of the self-energy at $\omega=0$ [\ref{DOS-Selfenergy-etas}(j)] is attributed to the failure of the CPA to reproduce a Fermi-liquid state\cite{kakehashi2004coherent}. However, this does not affect its conclusion regarding the metallic phase. Therefore, by conducting comprehensive analyses of the band gap, double occupancy, DOS, imaginary part of self-energies, as well as Lorentzian broadening factor dependence of DOS at the Fermi level, we distinguish a Mott insulating phase, a band insulating phase, two distinct site-selective phases, and a metallic phase within the paramagnetic phase diagrams of this twisted system, which have been summarized in Fig.\ref{phase-diagram-U}.

\begin{figure}[htbp]
\includegraphics[width=0.48\textwidth,height=0.32\textwidth]{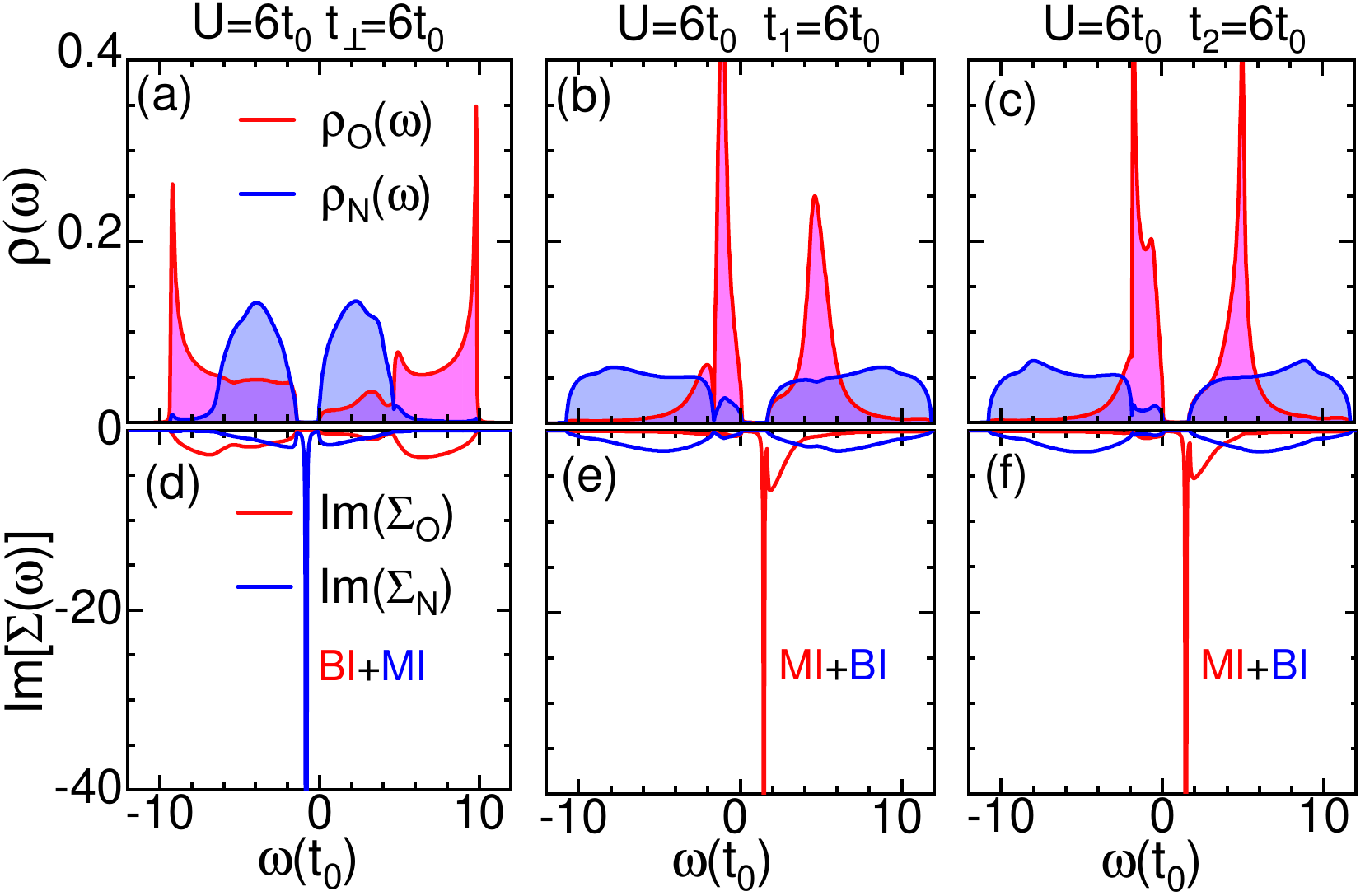}
\caption{The upper panel [(a)-(c)] illustrates the DOS at various parameter points, where $\rho_O(\omega)$ (pink shadow) and $\rho_N(\omega)$ (blue shadow) denote the DOS at each site in region O and region N of the supercell, respectively. The lower panel [(d)-(f)] presents the imaginary part of self-energies at various parameter points, where $\rm{Im}(\Sigma_O)$ (red line) and $\rm{Im}(\Sigma_N)$ (blue line) describe separately the imaginary part of self-energy at each site in region O and region N. (a) and (d) adopt the parameters of $U=6t_0$, $t_{\perp}=6t_0$, $t_1/t_0=t_2/t_0=0$, and $\Delta_O-\Delta_N=1.25t_0$. $U=6t_0$, $t_{1}=6t_0$, $t_\perp/t_0=t_2/t_0=0$, and $\Delta_O-\Delta_N=1.25t_0$ are used in (b) and (e). $U=6t_0$, $t_{2}=6t_0$, $t_\perp/t_0=t_1/t_0=0$, and $\Delta_O-\Delta_N=1.25t_0$ are employed in (c) and (f).}
\label{DOS-delta}
\end{figure}

Finally, we will examine the effect of an on-site potential difference on the stability of the site-selective insulating phases. As we know, the Mott insulating state has already been pointed out to be unstable to the ionic potential in the ionic Hubbard model\cite{garg2006can}. In the twisted system we studied, due to the emergence of two types of inequivalent sites, it also possesses two distinct on-site potentials including $\Delta_O$ and $\Delta_N$ which are analogous to the ionic potentials. Thus, it is necessary to study whether the site-selective insulating phases (the Mott insulating state present at some sites) disappear as long as there is an on-site potential difference. In Fig. \ref{DOS-delta}, we demonstrate the DOS and the imaginary part of the self-energy for the site-selective insulating phases (at three parameter points) under the effect of a moderate on-site potential difference with $\Delta_O-\Delta_N=1.25t_0$. It is
clear from Fig. \ref{DOS-delta}(a)-\ref{DOS-delta}(c) that, although the on-site potential difference breaks the particle-hole symmetry, the band gap still preserves, indicating the system in certain insulating phases. By further examining the imaginary parts of the self-energies of these insulating phases, we discover that $\rm{Im}(\Sigma_N)$ in Fig. \ref{DOS-delta}(d) as well as $\rm{Im}(\Sigma_O)$ in both Fig. \ref{DOS-delta}(e) and Fig. \ref{DOS-delta}(f) exhibit a divergent behavior within the band gap at a nonzero frequency. It has been pointed out that this divergence means the infinite values of both the scattering rate and the effective mass of quasiparticles, namely, the Mott physics at corresponding sites induced by strong electronic correlation\cite{xu2016gate}. Conversely, $\rm{Im}(\Sigma_O)$ in Fig. \ref{DOS-delta}(d) as well as $\rm{Im}(\Sigma_N)$ in both Fig. \ref{DOS-delta}(e) and Fig. \ref{DOS-delta}(f) are negligibly small near the Fermi level, an indication of a band insulating state at related sites. Therefore, these two site-selective insulating phases are still stable even in the presence of a moderate on-site potential difference.
\section{DISCUSSION}\label{DISCUSSION}
In this paper, we have demonstrated the presence of site-selective insulating phases in the half-filled Hubbard model on a twisted bilayer square lattice using the CPA. While the CPA provides a critical on-site Coulomb interaction of $U_c\approx{W/2}=4t_0$ ($W$ is the Bloch bandwidth) for the metal-Mott insulator transition in the monolayer square lattice, which differs slightly from other methods\cite{gull2013superconductivity,ruger2014phase} (consistent with Ref.\onlinecite{groeber1998paramagnetic}), previous research indicates that the critical $U_c$ is quite complicated and depends strongly on the Hubbard model under investigation and the methods employed\cite{luo2000higher}.
Regardless of the specific critical phase transition point, the CPA remains a reliable approach capable of handling phase transitions among band insulating, Mott insulating, and metallic states\cite{hoang2010metal,rowlands2014inclusion,rowlands2014disappearance,xu2016gate,xu2016interaction}. Therefore, our discovery of the site-selective insulating phases is qualitatively reliable, as the site-selective insulating phases we found are merely composed of a band insulating state and a Mott insulating state. Noticeably, the CPA ignores the spatial fluctuations of the effective medium (the self-energy is momentum-independent), it is interesting to employ other methods like CDMFT\cite{kotliar2001cellular} to precisely determine the detailed phase boundaries when taking into account the effects of short-range spatial correlations.

As we know, as long as a weak on-site Coulomb interaction is present, an antiferromagnetic insulating phase emerges in the region of weak interlayer hopping within the magnetic phase diagram for the Hubbard model on untwisted bilayer square lattice, which is attributed to the perfect nesting property of the Fermi surface within the noninteracting system\cite{golor2014ground,ruger2014phase}. However, when the twist is employed between two layers, the perfect nesting property of the Fermi surface is destroyed by the hoppings between two relatively twisted layers, which may have a significant impact on its phase diagram. Besides, we observe the site-selective phases even in the paramagnetic phase diagrams. Therefore, it is foreseeable that there will be various fascinating phases in the magnetic phase diagram of twisted bilayer square lattices.

Twist and pressure can manipulate interlayer hoppings and thereby induce novel phases in the layered systems, for example, the interlayer ferromagnetic and interlayer antiferromagnetic states can coexist in bilayer CrI$_3$ when twisting two layers with an angle of $\theta\leq3^{\circ}$\cite{xu2022coexisting}, the phase transition from a nonmagnetic state to a ferromagnetic state emerges in twisted bilayer graphene when applying pressure perpendicular to the layers\cite{pant2022phase}. We have demonstrated that the site-selective phases depend on the detailed values of interlayer hoppings. Therefore, it is quite intriguing to consider whether more complicated site-selective phases exist at different twisted angles or if phase transitions from the site-selective insulating phases to other phases occur under pressure.

Our discovery of the site-selective insulating phases in twisted bilayer square lattice suggests twist as an effective approach to access a site-selective phase in the strongly correlated system. While our proposal remains a theoretical prediction, it may still stimulate tremendous research interest for the following reasons. Firstly, one may be interested in exploring whether a site-selective superconducting phase, where certain sites exhibit a superconducting state while the others are in a normal state, exists in twisted layered superconducting materials since there are currently mature experimental techniques for synthesizing monolayer superconducting material\cite{yu2019high,zhao2019sign} and constructing two superconducting crystals along the $c$ axis with a twist ($c$-axis twisted Josephson junctions)\cite{li1999bi,takano2002d,latyshev2004c,yang2018pi,zhu2021presence,zhao2021emergent,lee2021twisted}. Secondly, while we predict the site-selective phase in a square lattice, it is possible that a site-selective phase may also be present in twisted systems with other lattice structures, making it intriguing to explore a site-selective insulating phase with the coexistence of band and Mott insulating states in twisted layered strongly correlated materials, not limited to square lattice materials. Finally, the nature of the Mott insulator observed in magic-angle twisted bilayer graphene\cite{cao2018correlated} is worth reexamining since the DOS of the flat bands is primarily contributed by the atoms located at the AA-stacking zone\cite{trambly2010localization}. Considering that, in the absence of Coulomb interactions, atoms in the AB-stacking zone lack electrons near the Fermi level, it is likely that this observed insulating phase (induced by the interaction) is a site-selective phase, where the atoms located at the AA-stacking zone exhibit a Mott insulating state while others maintain a band insulating state.

\section{conclusion}\label{Conclusion}
In conclusion, we systematically investigate the paramagnetic phase diagrams of the half-filled Hubbard model on a twisted bilayer square lattice by employing the CPA. The site-selective insulating phases are discovered, characterized by the coexistence of band insulating states at some sites while Mott insulating states at the remainings, in addition to the pure metallic, band insulating, and Mott insulating phases in the whole lattice. We attribute the appearance of site-selective insulating phases to the differentiation of interlayer hoppings in different regions with the help of strong on-site Coulomb repulsions. We find that the site-selective insulating phases are stable even in the presence of a moderate site-dependent on-site potential. Our findings not only demonstrate a fascinating phenomenon that a band insulating state can coexist with a Mott insulating state in strongly correlated systems but also suggest that varieties of site-selective phases might be realized by applying twist to layered materials.

\begin{center}
\textbf{ACKNOWLEDGEMENT}
\end{center}
This work is supported by National Natural Science Foundation of China (NOs.12004283, 12274324) and Shanghai Science and technology program (No.21JC405700).
\appendix
\begin{widetext}
\section{The mapping of models from a disordered alloy to an effective medium within the framework of the CPA}\label{mapping}
To derive the mapping from model \eqref{alloyModel} to model \eqref{CPAModel} within the framework of the CPA, we start by calculating the single-particle Green's function of the effective medium. Based on model \eqref{CPAModel}, the corresponding Hamiltonian of the effective medium in momentum space reads
\begin{equation}
H_{eff}=\sum_{k,\sigma} \big|{\psi_{\boldsymbol{k},\sigma}}\big\rangle{\widehat{\mathcal{M}}(\boldsymbol{k})}\big\langle{\psi_{\boldsymbol{k},\sigma}\big|},
\end{equation}
where
\begin{tiny}
\begin{align}
\widehat{\mathcal{M}}(\boldsymbol{k})=
\left[
\begin{smallmatrix}
\Delta_O-\mu+\Sigma_{A1}   
&-t_{0}  
&-t_{0}e^{-ik_x}  
&-t_{0}e^{-i(k_x+k_y)}  
&-t_{0}e^{-ik_y} 
&-t_{\perp}  
&0  
&0  
&0  
&0\\
-t_{0}   
&\Delta_N-\mu+\Sigma_{B1}  
&-t_{0}  
&-t_{0}e^{-ik_y}  
&-t_{0}  
&0  
&-t_{2}  
&0  
&0  
&-t_{1}\\
-t_{0}e^{ik_x}   
&-t_{0}  
&\Delta_N-\mu+\Sigma_{C1}  
&-t_{0}  
&-t_{0}e^{ik_x}  
&0  
&0  
&0  
&-t_{1}  
&-t_{2}\\
-t_{0}e^{i(k_x+k_y)}   
&-t_{0}e^{ik_y}  
&-t_{0}  
&\Delta_N-\mu+\Sigma_{D1}  
&-t_{0}  
&0  
&0  
&-t_{1}  
&-t_{2}  
&0\\
-t_{0}e^{ik_y}   
&-t_{0}  
&-t_{0}e^{-ik_x}  
&-t_{0}  
&\Delta_N-\mu+\Sigma_{E1}  
&0  
&-t_{1}  
&-t_{2}  
&0  
&0\\
-t_{\perp}   
&0  
&0  
&0  
&0  
&\Delta_O-\mu+\Sigma_{A2}  
&-t_{0}  
&-t_{0}e^{-ik_y}  
&-t_{0}e^{-i(k_x+k_y)}  
&-t_{0}e^{-ik_x}\\
0   
&-t_{2}  
&0  
&0  
&-t_{1}  
&-t_{0}  
&\Delta_N-\mu+\Sigma_{B2}  
&-t_{0}  
&-t_{0}e^{-ik_x}  
&-t_{0}\\
0   
&0  
&0  
&-t_{1}  
&-t_{2}  
&-t_{0}e^{ik_y}  
&-t_{0}  
&\Delta_N-\mu+\Sigma_{C2}  
&-t_{0}  
&-t_{0}e^{ik_y}\\
0   
&0  
&-t_{1}  
&-t_{2}  
&0  
&-t_{0}e^{i(k_x+k_y)}  
&-t_{0}e^{ik_x}  
&-t_{0}  
&\Delta_N-\mu+\Sigma_{D2}  
&-t_{0}\\
0   
&-t_{1}  
&-t_{2}  
&0  
&0  
&-t_{0}e^{ik_x}  
&-t_{0}  
&-t_{0}e^{-ik_y}  
&-t_{0}  
&\Delta_N-\mu+\Sigma_{E2}  
\end{smallmatrix}
\right]
\label{matrixM}
\end{align}
\end{tiny}
and
\begin{align}
\big|{\psi_{\boldsymbol{k},\sigma}}\big\rangle
=
\left(
\begin{matrix}
C_{kA1\sigma}^{\dag},&
C_{kB1\sigma}^{\dag},&
C_{kC1\sigma}^{\dag},&
C_{kD1\sigma}^{\dag},&
C_{kE1\sigma}^{\dag},&
C_{kA2\sigma}^{\dag},&
C_{kB2\sigma}^{\dag},&
C_{kC2\sigma}^{\dag},&
C_{kD2\sigma}^{\dag},&
C_{kE2\sigma}^{\dag}
\end{matrix}
\right)
\end{align}
Here, all of these self-energies $\Sigma_{A1},\cdots,\Sigma_{E2}$ are both complex and energy-dependent. Noticeably, the Hamiltonian matrix $\widehat{\mathcal{M}}(\boldsymbol{k})$ omits the spin indices as we are interested in the paramagnetic phase. Thus, Green's function of the effective medium in momentum space can be readily calculated as
\begin{equation}
G^{eff}(\boldsymbol{k},\omega)=\frac{1}{\omega-\widehat{\mathcal{M}}(\boldsymbol{k})+i\eta},
\label{effGK}
\end{equation}
where $\eta$ stands for the Lorentzian broadening factor. Using Green's function in momentum space, the corresponding Green's function of the effective medium in real space reads
\begin{equation}
G^{eff}_{ism,ism}(\omega)=\frac{1}{\Omega_{BZ}}\int_{\Omega_{BZ}} G^{eff}_{sm,sm}(\boldsymbol{k},\omega) d\boldsymbol{k},
\label{effGR}
\end{equation}
where the integral is over the first Brillouin zone of the system. Then, the cavity Green's function $\mathcal{G}_{ism}(\omega)$ can be obtained through the Dyson equation
\begin{equation}
\mathcal{G}_{ism}^{-1}(\omega)=\big[G^{eff}_{ism,ism}(\omega)\big]^{-1}+\Sigma_{sm}(\omega)
\label{cavity}
\end{equation}
for a given $s$ sublattice at $m$ layer of $i$th supercell, which describes a medium with removed self-energy at a chosen site. It is necessary to mention that the self-energies of the effective medium arise from the disordered potentials of the disordered alloy within the framework of the CPA, suggesting that the cavity Green's function for a given site of the effective medium is equal to that of the disordered alloy. Therefore, the cavity can now be filled by a real ``impurity" with disorder potential, resulting in an impurity Green's function of the disordered alloy
\begin{equation}
G_{ism}(\omega)=\frac{1}{\mathcal {G}_{ism}^{-1}(\omega)-E_{ism}}
\end{equation}
with impurity configurations of
\begin{equation}
\left\{
\begin{aligned}
E_{ism}&=0,\quad P_0=1-\langle{n_{ism\bar{\sigma}}}\rangle\\
E_{ism}&=U,\quad P_U=\langle{n_{ism\bar{\sigma}}}\rangle
\end{aligned}
\right.
.\label{probability}
\end{equation}
Then, the average Green's function of the disordered alloy can be calculated by summing all the impurity Green's functions with corresponding probability weights, namely
\begin{equation}
\langle G_{ism}(\omega)\rangle =\frac{P_0}{\mathcal {G}_{ism}^{-1}(\omega)-0}+\frac{P_U}{\mathcal {G}_{ism}^{-1}(\omega)-U},
\label{ave-Green}
\end{equation}
Once the average Green's function of the disordered alloy and the Green's function of the effective medium satisfies
\begin{equation}
\langle G_{ism}(\omega)\rangle =G^{eff}_{ism,ism}(\omega),
\label{CPA}
\end{equation}
the model \eqref{alloyModel} can be successfully mapped into model \eqref{CPAModel}. Noticeably, since we focus on the case at half filling, the extra condition must be satisfied
\begin{equation}
\sum\limits_{sm}\langle{n_{ism\sigma}}\rangle=5
\end{equation}
where
\begin{equation}
\langle n_{ism\sigma}\rangle=-\frac{1}{\pi}\int_{-\infty}^{0}\text{Im} \Big[G^{eff}_{ism,ism}(\omega)\Big] d\omega.
\end{equation}
These calculated average occupation number need to be applied to compute new probability of impurities \eqref{probability}.
\end{widetext}
\bibliography{CPA_reference}

\providecommand{\noopsort}[1]{}\providecommand{\singleletter}[1]{#1}%
\begin{thebibliography}{89}%
\makeatletter
\providecommand \@ifxundefined [1]{%
 \@ifx{#1\undefined}
}%
\providecommand \@ifnum [1]{%
 \ifnum #1\expandafter \@firstoftwo
 \else \expandafter \@secondoftwo
 \fi
}%
\providecommand \@ifx [1]{%
 \ifx #1\expandafter \@firstoftwo
 \else \expandafter \@secondoftwo
 \fi
}%
\providecommand \natexlab [1]{#1}%
\providecommand \enquote  [1]{``#1''}%
\providecommand \bibnamefont  [1]{#1}%
\providecommand \bibfnamefont [1]{#1}%
\providecommand \citenamefont [1]{#1}%
\providecommand \href@noop [0]{\@secondoftwo}%
\providecommand \href [0]{\begingroup \@sanitize@url \@href}%
\providecommand \@href[1]{\@@startlink{#1}\@@href}%
\providecommand \@@href[1]{\endgroup#1\@@endlink}%
\providecommand \@sanitize@url [0]{\catcode `\\12\catcode `\$12\catcode
  `\&12\catcode `\#12\catcode `\^12\catcode `\_12\catcode `\%12\relax}%
\providecommand \@@startlink[1]{}%
\providecommand \@@endlink[0]{}%
\providecommand \url  [0]{\begingroup\@sanitize@url \@url }%
\providecommand \@url [1]{\endgroup\@href {#1}{\urlprefix }}%
\providecommand \urlprefix  [0]{URL }%
\providecommand \Eprint [0]{\href }%
\providecommand \doibase [0]{http://dx.doi.org/}%
\providecommand \selectlanguage [0]{\@gobble}%
\providecommand \bibinfo  [0]{\@secondoftwo}%
\providecommand \bibfield  [0]{\@secondoftwo}%
\providecommand \translation [1]{[#1]}%
\providecommand \BibitemOpen [0]{}%
\providecommand \bibitemStop [0]{}%
\providecommand \bibitemNoStop [0]{.\EOS\space}%
\providecommand \EOS [0]{\spacefactor3000\relax}%
\providecommand \BibitemShut  [1]{\csname bibitem#1\endcsname}%
\let\auto@bib@innerbib\@empty
\bibitem [{\citenamefont {Cao}\ \emph {et~al.}(2018{\natexlab{a}})\citenamefont
  {Cao}, \citenamefont {Fatemi}, \citenamefont {Demir}, \citenamefont {Fang},
  \citenamefont {Tomarken}, \citenamefont {Luo}, \citenamefont
  {Sanchez-Yamagishi}, \citenamefont {Watanabe}, \citenamefont {Taniguchi},
  \citenamefont {Kaxiras} \emph {et~al.}}]{cao2018correlated}%
  \BibitemOpen
  \bibfield  {author} {\bibinfo {author} {\bibfnamefont {Yuan}\ \bibnamefont
  {Cao}}, \bibinfo {author} {\bibfnamefont {Valla}\ \bibnamefont {Fatemi}},
  \bibinfo {author} {\bibfnamefont {Ahmet}\ \bibnamefont {Demir}}, \bibinfo
  {author} {\bibfnamefont {Shiang}\ \bibnamefont {Fang}}, \bibinfo {author}
  {\bibfnamefont {Spencer~L}\ \bibnamefont {Tomarken}}, \bibinfo {author}
  {\bibfnamefont {Jason~Y}\ \bibnamefont {Luo}}, \bibinfo {author}
  {\bibfnamefont {Javier~D}\ \bibnamefont {Sanchez-Yamagishi}}, \bibinfo
  {author} {\bibfnamefont {Kenji}\ \bibnamefont {Watanabe}}, \bibinfo {author}
  {\bibfnamefont {Takashi}\ \bibnamefont {Taniguchi}}, \bibinfo {author}
  {\bibfnamefont {Efthimios}\ \bibnamefont {Kaxiras}},  \emph {et~al.},\
  }\bibfield  {title} {\enquote {\bibinfo {title} {{Correlated insulator
  behaviour at half-filling in magic-angle graphene superlattices}},}\
  }\href@noop {} {\bibfield  {journal} {\bibinfo  {journal} {Nature}\ }\textbf
  {\bibinfo {volume} {556}},\ \bibinfo {pages} {80--84} (\bibinfo {year}
  {2018}{\natexlab{a}})}\BibitemShut {NoStop}%
\bibitem [{\citenamefont {Chen}\ \emph
  {et~al.}(2019{\natexlab{a}})\citenamefont {Chen}, \citenamefont {Jiang},
  \citenamefont {Wu}, \citenamefont {Lyu}, \citenamefont {Li}, \citenamefont
  {Chittari}, \citenamefont {Watanabe}, \citenamefont {Taniguchi},
  \citenamefont {Shi}, \citenamefont {Jung} \emph {et~al.}}]{chen2019evidence}%
  \BibitemOpen
  \bibfield  {author} {\bibinfo {author} {\bibfnamefont {Guorui}\ \bibnamefont
  {Chen}}, \bibinfo {author} {\bibfnamefont {Lili}\ \bibnamefont {Jiang}},
  \bibinfo {author} {\bibfnamefont {Shuang}\ \bibnamefont {Wu}}, \bibinfo
  {author} {\bibfnamefont {Bosai}\ \bibnamefont {Lyu}}, \bibinfo {author}
  {\bibfnamefont {Hongyuan}\ \bibnamefont {Li}}, \bibinfo {author}
  {\bibfnamefont {Bheema~Lingam}\ \bibnamefont {Chittari}}, \bibinfo {author}
  {\bibfnamefont {Kenji}\ \bibnamefont {Watanabe}}, \bibinfo {author}
  {\bibfnamefont {Takashi}\ \bibnamefont {Taniguchi}}, \bibinfo {author}
  {\bibfnamefont {Zhiwen}\ \bibnamefont {Shi}}, \bibinfo {author}
  {\bibfnamefont {Jeil}\ \bibnamefont {Jung}},  \emph {et~al.},\ }\bibfield
  {title} {\enquote {\bibinfo {title} {{Evidence of a gate-tunable Mott
  insulator in a trilayer graphene moir{\'e} superlattice}},}\ }\href@noop {}
  {\bibfield  {journal} {\bibinfo  {journal} {Nature Physics}\ }\textbf
  {\bibinfo {volume} {15}},\ \bibinfo {pages} {237--241} (\bibinfo {year}
  {2019}{\natexlab{a}})}\BibitemShut {NoStop}%
\bibitem [{\citenamefont {Regan}\ \emph {et~al.}(2020)\citenamefont {Regan},
  \citenamefont {Wang}, \citenamefont {Jin}, \citenamefont {Bakti~Utama},
  \citenamefont {Gao}, \citenamefont {Wei}, \citenamefont {Zhao}, \citenamefont
  {Zhao}, \citenamefont {Zhang}, \citenamefont {Yumigeta} \emph
  {et~al.}}]{regan2020mott}%
  \BibitemOpen
  \bibfield  {author} {\bibinfo {author} {\bibfnamefont {Emma~C}\ \bibnamefont
  {Regan}}, \bibinfo {author} {\bibfnamefont {Danqing}\ \bibnamefont {Wang}},
  \bibinfo {author} {\bibfnamefont {Chenhao}\ \bibnamefont {Jin}}, \bibinfo
  {author} {\bibfnamefont {M~Iqbal}\ \bibnamefont {Bakti~Utama}}, \bibinfo
  {author} {\bibfnamefont {Beini}\ \bibnamefont {Gao}}, \bibinfo {author}
  {\bibfnamefont {Xin}\ \bibnamefont {Wei}}, \bibinfo {author} {\bibfnamefont
  {Sihan}\ \bibnamefont {Zhao}}, \bibinfo {author} {\bibfnamefont {Wenyu}\
  \bibnamefont {Zhao}}, \bibinfo {author} {\bibfnamefont {Zuocheng}\
  \bibnamefont {Zhang}}, \bibinfo {author} {\bibfnamefont {Kentaro}\
  \bibnamefont {Yumigeta}},  \emph {et~al.},\ }\bibfield  {title} {\enquote
  {\bibinfo {title} {{Mott and generalized Wigner crystal states in WSe2/WS2
  moir{\'e} superlattices}},}\ }\href@noop {} {\bibfield  {journal} {\bibinfo
  {journal} {Nature}\ }\textbf {\bibinfo {volume} {579}},\ \bibinfo {pages}
  {359--363} (\bibinfo {year} {2020})}\BibitemShut {NoStop}%
\bibitem [{\citenamefont {Tang}\ \emph {et~al.}(2020)\citenamefont {Tang},
  \citenamefont {Li}, \citenamefont {Li}, \citenamefont {Xu}, \citenamefont
  {Liu}, \citenamefont {Barmak}, \citenamefont {Watanabe}, \citenamefont
  {Taniguchi}, \citenamefont {MacDonald}, \citenamefont {Shan} \emph
  {et~al.}}]{tang2020simulation}%
  \BibitemOpen
  \bibfield  {author} {\bibinfo {author} {\bibfnamefont {Yanhao}\ \bibnamefont
  {Tang}}, \bibinfo {author} {\bibfnamefont {Lizhong}\ \bibnamefont {Li}},
  \bibinfo {author} {\bibfnamefont {Tingxin}\ \bibnamefont {Li}}, \bibinfo
  {author} {\bibfnamefont {Yang}\ \bibnamefont {Xu}}, \bibinfo {author}
  {\bibfnamefont {Song}\ \bibnamefont {Liu}}, \bibinfo {author} {\bibfnamefont
  {Katayun}\ \bibnamefont {Barmak}}, \bibinfo {author} {\bibfnamefont {Kenji}\
  \bibnamefont {Watanabe}}, \bibinfo {author} {\bibfnamefont {Takashi}\
  \bibnamefont {Taniguchi}}, \bibinfo {author} {\bibfnamefont {Allan~H}\
  \bibnamefont {MacDonald}}, \bibinfo {author} {\bibfnamefont {Jie}\
  \bibnamefont {Shan}},  \emph {et~al.},\ }\bibfield  {title} {\enquote
  {\bibinfo {title} {{Simulation of Hubbard model physics in WSe2/WS2 moir{\'e}
  superlattices}},}\ }\href@noop {} {\bibfield  {journal} {\bibinfo  {journal}
  {Nature}\ }\textbf {\bibinfo {volume} {579}},\ \bibinfo {pages} {353--358}
  (\bibinfo {year} {2020})}\BibitemShut {NoStop}%
\bibitem [{\citenamefont {Li}\ \emph {et~al.}(2021)\citenamefont {Li},
  \citenamefont {Jiang}, \citenamefont {Li}, \citenamefont {Zhang},
  \citenamefont {Kang}, \citenamefont {Zhu}, \citenamefont {Watanabe},
  \citenamefont {Taniguchi}, \citenamefont {Chowdhury}, \citenamefont {Fu}
  \emph {et~al.}}]{li2021continuous}%
  \BibitemOpen
  \bibfield  {author} {\bibinfo {author} {\bibfnamefont {Tingxin}\ \bibnamefont
  {Li}}, \bibinfo {author} {\bibfnamefont {Shengwei}\ \bibnamefont {Jiang}},
  \bibinfo {author} {\bibfnamefont {Lizhong}\ \bibnamefont {Li}}, \bibinfo
  {author} {\bibfnamefont {Yang}\ \bibnamefont {Zhang}}, \bibinfo {author}
  {\bibfnamefont {Kaifei}\ \bibnamefont {Kang}}, \bibinfo {author}
  {\bibfnamefont {Jiacheng}\ \bibnamefont {Zhu}}, \bibinfo {author}
  {\bibfnamefont {Kenji}\ \bibnamefont {Watanabe}}, \bibinfo {author}
  {\bibfnamefont {Takashi}\ \bibnamefont {Taniguchi}}, \bibinfo {author}
  {\bibfnamefont {Debanjan}\ \bibnamefont {Chowdhury}}, \bibinfo {author}
  {\bibfnamefont {Liang}\ \bibnamefont {Fu}},  \emph {et~al.},\ }\bibfield
  {title} {\enquote {\bibinfo {title} {{Continuous Mott transition in
  semiconductor moir{\'e} superlattices}},}\ }\href@noop {} {\bibfield
  {journal} {\bibinfo  {journal} {Nature}\ }\textbf {\bibinfo {volume} {597}},\
  \bibinfo {pages} {350--354} (\bibinfo {year} {2021})}\BibitemShut {NoStop}%
\bibitem [{\citenamefont {Cao}\ \emph {et~al.}(2018{\natexlab{b}})\citenamefont
  {Cao}, \citenamefont {Fatemi}, \citenamefont {Fang}, \citenamefont
  {Watanabe}, \citenamefont {Taniguchi}, \citenamefont {Kaxiras},\ and\
  \citenamefont {Jarillo-Herrero}}]{cao2018unconventional}%
  \BibitemOpen
  \bibfield  {author} {\bibinfo {author} {\bibfnamefont {Yuan}\ \bibnamefont
  {Cao}}, \bibinfo {author} {\bibfnamefont {Valla}\ \bibnamefont {Fatemi}},
  \bibinfo {author} {\bibfnamefont {Shiang}\ \bibnamefont {Fang}}, \bibinfo
  {author} {\bibfnamefont {Kenji}\ \bibnamefont {Watanabe}}, \bibinfo {author}
  {\bibfnamefont {Takashi}\ \bibnamefont {Taniguchi}}, \bibinfo {author}
  {\bibfnamefont {Efthimios}\ \bibnamefont {Kaxiras}}, \ and\ \bibinfo {author}
  {\bibfnamefont {Pablo}\ \bibnamefont {Jarillo-Herrero}},\ }\bibfield  {title}
  {\enquote {\bibinfo {title} {{Unconventional superconductivity in magic-angle
  graphene superlattices}},}\ }\href@noop {} {\bibfield  {journal} {\bibinfo
  {journal} {Nature}\ }\textbf {\bibinfo {volume} {556}},\ \bibinfo {pages}
  {43--50} (\bibinfo {year} {2018}{\natexlab{b}})}\BibitemShut {NoStop}%
\bibitem [{\citenamefont {Chen}\ \emph
  {et~al.}(2019{\natexlab{b}})\citenamefont {Chen}, \citenamefont {Sharpe},
  \citenamefont {Gallagher}, \citenamefont {Rosen}, \citenamefont {Fox},
  \citenamefont {Jiang}, \citenamefont {Lyu}, \citenamefont {Li}, \citenamefont
  {Watanabe}, \citenamefont {Taniguchi} \emph {et~al.}}]{chen2019signatures}%
  \BibitemOpen
  \bibfield  {author} {\bibinfo {author} {\bibfnamefont {Guorui}\ \bibnamefont
  {Chen}}, \bibinfo {author} {\bibfnamefont {Aaron~L}\ \bibnamefont {Sharpe}},
  \bibinfo {author} {\bibfnamefont {Patrick}\ \bibnamefont {Gallagher}},
  \bibinfo {author} {\bibfnamefont {Ilan~T}\ \bibnamefont {Rosen}}, \bibinfo
  {author} {\bibfnamefont {Eli~J}\ \bibnamefont {Fox}}, \bibinfo {author}
  {\bibfnamefont {Lili}\ \bibnamefont {Jiang}}, \bibinfo {author}
  {\bibfnamefont {Bosai}\ \bibnamefont {Lyu}}, \bibinfo {author} {\bibfnamefont
  {Hongyuan}\ \bibnamefont {Li}}, \bibinfo {author} {\bibfnamefont {Kenji}\
  \bibnamefont {Watanabe}}, \bibinfo {author} {\bibfnamefont {Takashi}\
  \bibnamefont {Taniguchi}},  \emph {et~al.},\ }\bibfield  {title} {\enquote
  {\bibinfo {title} {{Signatures of tunable superconductivity in a trilayer
  graphene moir{\'e} superlattice}},}\ }\href@noop {} {\bibfield  {journal}
  {\bibinfo  {journal} {Nature}\ }\textbf {\bibinfo {volume} {572}},\ \bibinfo
  {pages} {215--219} (\bibinfo {year} {2019}{\natexlab{b}})}\BibitemShut
  {NoStop}%
\bibitem [{\citenamefont {Park}\ \emph {et~al.}(2019)\citenamefont {Park},
  \citenamefont {Kim}, \citenamefont {Cho},\ and\ \citenamefont
  {Lee}}]{park2019higher}%
  \BibitemOpen
  \bibfield  {author} {\bibinfo {author} {\bibfnamefont {Moon~Jip}\
  \bibnamefont {Park}}, \bibinfo {author} {\bibfnamefont {Youngkuk}\
  \bibnamefont {Kim}}, \bibinfo {author} {\bibfnamefont {Gil~Young}\
  \bibnamefont {Cho}}, \ and\ \bibinfo {author} {\bibfnamefont {SungBin}\
  \bibnamefont {Lee}},\ }\bibfield  {title} {\enquote {\bibinfo {title}
  {{Higher-order topological insulator in twisted bilayer graphene}},}\
  }\href@noop {} {\bibfield  {journal} {\bibinfo  {journal} {Physical review
  letters}\ }\textbf {\bibinfo {volume} {123}},\ \bibinfo {pages} {216803}
  (\bibinfo {year} {2019})}\BibitemShut {NoStop}%
\bibitem [{\citenamefont {Can}\ \emph {et~al.}(2021)\citenamefont {Can},
  \citenamefont {Tummuru}, \citenamefont {Day}, \citenamefont {Elfimov},
  \citenamefont {Damascelli},\ and\ \citenamefont {Franz}}]{can2021high}%
  \BibitemOpen
  \bibfield  {author} {\bibinfo {author} {\bibfnamefont {Oguzhan}\ \bibnamefont
  {Can}}, \bibinfo {author} {\bibfnamefont {Tarun}\ \bibnamefont {Tummuru}},
  \bibinfo {author} {\bibfnamefont {Ryan~P}\ \bibnamefont {Day}}, \bibinfo
  {author} {\bibfnamefont {Ilya}\ \bibnamefont {Elfimov}}, \bibinfo {author}
  {\bibfnamefont {Andrea}\ \bibnamefont {Damascelli}}, \ and\ \bibinfo {author}
  {\bibfnamefont {Marcel}\ \bibnamefont {Franz}},\ }\bibfield  {title}
  {\enquote {\bibinfo {title} {{High-temperature topological superconductivity
  in twisted double-layer copper oxides}},}\ }\href@noop {} {\bibfield
  {journal} {\bibinfo  {journal} {Nature Physics}\ }\textbf {\bibinfo {volume}
  {17}},\ \bibinfo {pages} {519--524} (\bibinfo {year} {2021})}\BibitemShut
  {NoStop}%
\bibitem [{\citenamefont {Eugenio}\ and\ \citenamefont
  {Vafek}(2023)}]{eugenio2023twisted}%
  \BibitemOpen
  \bibfield  {author} {\bibinfo {author} {\bibfnamefont {Paul~Myles}\
  \bibnamefont {Eugenio}}\ and\ \bibinfo {author} {\bibfnamefont {Oskar}\
  \bibnamefont {Vafek}},\ }\bibfield  {title} {\enquote {\bibinfo {title}
  {{Twisted-bilayer FeSe and the Fe-based superlattices}},}\ }\href@noop {}
  {\bibfield  {journal} {\bibinfo  {journal} {SciPost Physics}\ }\textbf
  {\bibinfo {volume} {15}},\ \bibinfo {pages} {081} (\bibinfo {year}
  {2023})}\BibitemShut {NoStop}%
\bibitem [{\citenamefont {Hubbard}(1964)}]{hubbard1964electron3}%
  \BibitemOpen
  \bibfield  {author} {\bibinfo {author} {\bibfnamefont {John}\ \bibnamefont
  {Hubbard}},\ }\bibfield  {title} {\enquote {\bibinfo {title} {{Electron
  correlations in narrow energy bands III. An improved solution}},}\
  }\href@noop {} {\bibfield  {journal} {\bibinfo  {journal} {Proceedings of the
  Royal Society of London. Series A. Mathematical and Physical Sciences}\
  }\textbf {\bibinfo {volume} {281}},\ \bibinfo {pages} {401--419} (\bibinfo
  {year} {1964})}\BibitemShut {NoStop}%
\bibitem [{\citenamefont {Koga}\ \emph {et~al.}(2004)\citenamefont {Koga},
  \citenamefont {Kawakami}, \citenamefont {Rice},\ and\ \citenamefont
  {Sigrist}}]{koga2004orbital}%
  \BibitemOpen
  \bibfield  {author} {\bibinfo {author} {\bibfnamefont {Akihisa}\ \bibnamefont
  {Koga}}, \bibinfo {author} {\bibfnamefont {Norio}\ \bibnamefont {Kawakami}},
  \bibinfo {author} {\bibfnamefont {TM}~\bibnamefont {Rice}}, \ and\ \bibinfo
  {author} {\bibfnamefont {Manfred}\ \bibnamefont {Sigrist}},\ }\bibfield
  {title} {\enquote {\bibinfo {title} {{Orbital-selective Mott transitions in
  the degenerate Hubbard model}},}\ }\href@noop {} {\bibfield  {journal}
  {\bibinfo  {journal} {Physical review letters}\ }\textbf {\bibinfo {volume}
  {92}},\ \bibinfo {pages} {216402} (\bibinfo {year} {2004})}\BibitemShut
  {NoStop}%
\bibitem [{\citenamefont {de’Medici}\ \emph {et~al.}(2009)\citenamefont
  {de’Medici}, \citenamefont {Hassan}, \citenamefont {Capone},\ and\
  \citenamefont {Dai}}]{de2009orbital}%
  \BibitemOpen
  \bibfield  {author} {\bibinfo {author} {\bibfnamefont {Luca}\ \bibnamefont
  {de’Medici}}, \bibinfo {author} {\bibfnamefont {Syed~R}\ \bibnamefont
  {Hassan}}, \bibinfo {author} {\bibfnamefont {Massimo}\ \bibnamefont
  {Capone}}, \ and\ \bibinfo {author} {\bibfnamefont {Xi}~\bibnamefont {Dai}},\
  }\bibfield  {title} {\enquote {\bibinfo {title} {{Orbital-selective Mott
  transition out of band degeneracy lifting}},}\ }\href@noop {} {\bibfield
  {journal} {\bibinfo  {journal} {Physical review letters}\ }\textbf {\bibinfo
  {volume} {102}},\ \bibinfo {pages} {126401} (\bibinfo {year}
  {2009})}\BibitemShut {NoStop}%
\bibitem [{\citenamefont {Werner}\ and\ \citenamefont
  {Millis}(2007)}]{werner2007high}%
  \BibitemOpen
  \bibfield  {author} {\bibinfo {author} {\bibfnamefont {Philipp}\ \bibnamefont
  {Werner}}\ and\ \bibinfo {author} {\bibfnamefont {Andrew~J}\ \bibnamefont
  {Millis}},\ }\bibfield  {title} {\enquote {\bibinfo {title} {{High-spin to
  low-spin and orbital polarization transitions in multiorbital Mott
  systems}},}\ }\href@noop {} {\bibfield  {journal} {\bibinfo  {journal}
  {Physical review letters}\ }\textbf {\bibinfo {volume} {99}},\ \bibinfo
  {pages} {126405} (\bibinfo {year} {2007})}\BibitemShut {NoStop}%
\bibitem [{\citenamefont {Song}\ \emph {et~al.}(2015)\citenamefont {Song},
  \citenamefont {Lee},\ and\ \citenamefont {Zhang}}]{song2015possible}%
  \BibitemOpen
  \bibfield  {author} {\bibinfo {author} {\bibfnamefont {Ze-Yi}\ \bibnamefont
  {Song}}, \bibinfo {author} {\bibfnamefont {Hunpyo}\ \bibnamefont {Lee}}, \
  and\ \bibinfo {author} {\bibfnamefont {Yu-Zhong}\ \bibnamefont {Zhang}},\
  }\bibfield  {title} {\enquote {\bibinfo {title} {{Possible origin of orbital
  selective Mott transitions in iron-based superconductors and
  Ca$\rm{_{2-x}}$Sr$\rm{_{x}}$RuO$\rm{_{4}}$}},}\ }\href@noop {} {\bibfield
  {journal} {\bibinfo  {journal} {New Journal of Physics}\ }\textbf {\bibinfo
  {volume} {17}},\ \bibinfo {pages} {033034} (\bibinfo {year}
  {2015})}\BibitemShut {NoStop}%
\bibitem [{\citenamefont {Zhang}(2004)}]{zhang2004dimerization}%
  \BibitemOpen
  \bibfield  {author} {\bibinfo {author} {\bibfnamefont {YZ}~\bibnamefont
  {Zhang}},\ }\bibfield  {title} {\enquote {\bibinfo {title} {{Dimerization in
  a half-filled one-dimensional extended Hubbard model}},}\ }\href@noop {}
  {\bibfield  {journal} {\bibinfo  {journal} {Physical review letters}\
  }\textbf {\bibinfo {volume} {92}},\ \bibinfo {pages} {246404} (\bibinfo
  {year} {2004})}\BibitemShut {NoStop}%
\bibitem [{\citenamefont {Kancharla}\ and\ \citenamefont
  {Dagotto}(2007)}]{kancharla2007correlated}%
  \BibitemOpen
  \bibfield  {author} {\bibinfo {author} {\bibfnamefont
  {Srivenkateswara~Sarma}\ \bibnamefont {Kancharla}}\ and\ \bibinfo {author}
  {\bibfnamefont {E}~\bibnamefont {Dagotto}},\ }\bibfield  {title} {\enquote
  {\bibinfo {title} {{Correlated insulated phase suggests bond order between
  band and Mott insulators in two dimensions}},}\ }\href@noop {} {\bibfield
  {journal} {\bibinfo  {journal} {Physical review letters}\ }\textbf {\bibinfo
  {volume} {98}},\ \bibinfo {pages} {016402} (\bibinfo {year}
  {2007})}\BibitemShut {NoStop}%
\bibitem [{\citenamefont {Schulz}(1987)}]{schulz1987superconductivity}%
  \BibitemOpen
  \bibfield  {author} {\bibinfo {author} {\bibfnamefont {HJ}~\bibnamefont
  {Schulz}},\ }\bibfield  {title} {\enquote {\bibinfo {title}
  {{Superconductivity and antiferromagnetism in the two-dimensional Hubbard
  model: Scaling theory}},}\ }\href@noop {} {\bibfield  {journal} {\bibinfo
  {journal} {Europhysics Letters}\ }\textbf {\bibinfo {volume} {4}},\ \bibinfo
  {pages} {609} (\bibinfo {year} {1987})}\BibitemShut {NoStop}%
\bibitem [{\citenamefont {Maier}\ \emph {et~al.}(2005)\citenamefont {Maier},
  \citenamefont {Jarrell}, \citenamefont {Schulthess}, \citenamefont {Kent},\
  and\ \citenamefont {White}}]{maier2005systematic}%
  \BibitemOpen
  \bibfield  {author} {\bibinfo {author} {\bibfnamefont {Th~A}\ \bibnamefont
  {Maier}}, \bibinfo {author} {\bibfnamefont {M}~\bibnamefont {Jarrell}},
  \bibinfo {author} {\bibfnamefont {TC}~\bibnamefont {Schulthess}}, \bibinfo
  {author} {\bibfnamefont {PRC}\ \bibnamefont {Kent}}, \ and\ \bibinfo {author}
  {\bibfnamefont {JB}~\bibnamefont {White}},\ }\bibfield  {title} {\enquote
  {\bibinfo {title} {{Systematic study of d-wave superconductivity in the 2D
  repulsive Hubbard model}},}\ }\href@noop {} {\bibfield  {journal} {\bibinfo
  {journal} {Physical review letters}\ }\textbf {\bibinfo {volume} {95}},\
  \bibinfo {pages} {237001} (\bibinfo {year} {2005})}\BibitemShut {NoStop}%
\bibitem [{\citenamefont {Ogawa}\ \emph {et~al.}(1975)\citenamefont {Ogawa},
  \citenamefont {Kanda},\ and\ \citenamefont
  {Matsubara}}]{ogawa1975gutzwiller}%
  \BibitemOpen
  \bibfield  {author} {\bibinfo {author} {\bibfnamefont {Tohru}\ \bibnamefont
  {Ogawa}}, \bibinfo {author} {\bibfnamefont {Kunihiko}\ \bibnamefont {Kanda}},
  \ and\ \bibinfo {author} {\bibfnamefont {Takeo}\ \bibnamefont {Matsubara}},\
  }\bibfield  {title} {\enquote {\bibinfo {title} {{Gutzwiller approximation
  for antiferromagnetism in hubbard model}},}\ }\href@noop {} {\bibfield
  {journal} {\bibinfo  {journal} {Progress of Theoretical Physics}\ }\textbf
  {\bibinfo {volume} {53}},\ \bibinfo {pages} {614--633} (\bibinfo {year}
  {1975})}\BibitemShut {NoStop}%
\bibitem [{\citenamefont {Schulz}(1990)}]{schulz1990incommensurate}%
  \BibitemOpen
  \bibfield  {author} {\bibinfo {author} {\bibfnamefont {HJ}~\bibnamefont
  {Schulz}},\ }\bibfield  {title} {\enquote {\bibinfo {title} {{Incommensurate
  antiferromagnetism in the two-dimensional Hubbard model}},}\ }\href@noop {}
  {\bibfield  {journal} {\bibinfo  {journal} {Physical review letters}\
  }\textbf {\bibinfo {volume} {64}},\ \bibinfo {pages} {1445} (\bibinfo {year}
  {1990})}\BibitemShut {NoStop}%
\bibitem [{\citenamefont {Hirsch}\ and\ \citenamefont
  {Tang}(1989)}]{hirsch1989antiferromagnetism}%
  \BibitemOpen
  \bibfield  {author} {\bibinfo {author} {\bibfnamefont {JE}~\bibnamefont
  {Hirsch}}\ and\ \bibinfo {author} {\bibfnamefont {S}~\bibnamefont {Tang}},\
  }\bibfield  {title} {\enquote {\bibinfo {title} {{Antiferromagnetism in the
  two-dimensional Hubbard model}},}\ }\href@noop {} {\bibfield  {journal}
  {\bibinfo  {journal} {Physical review letters}\ }\textbf {\bibinfo {volume}
  {62}},\ \bibinfo {pages} {591} (\bibinfo {year} {1989})}\BibitemShut
  {NoStop}%
\bibitem [{\citenamefont {Wu}\ \emph {et~al.}(2018)\citenamefont {Wu},
  \citenamefont {Lovorn}, \citenamefont {Tutuc},\ and\ \citenamefont
  {MacDonald}}]{wu2018hubbard}%
  \BibitemOpen
  \bibfield  {author} {\bibinfo {author} {\bibfnamefont {Fengcheng}\
  \bibnamefont {Wu}}, \bibinfo {author} {\bibfnamefont {Timothy}\ \bibnamefont
  {Lovorn}}, \bibinfo {author} {\bibfnamefont {Emanuel}\ \bibnamefont {Tutuc}},
  \ and\ \bibinfo {author} {\bibfnamefont {Allan~H}\ \bibnamefont
  {MacDonald}},\ }\bibfield  {title} {\enquote {\bibinfo {title} {{Hubbard
  model physics in transition metal dichalcogenide moir{\'e} bands}},}\
  }\href@noop {} {\bibfield  {journal} {\bibinfo  {journal} {Physical review
  letters}\ }\textbf {\bibinfo {volume} {121}},\ \bibinfo {pages} {026402}
  (\bibinfo {year} {2018})}\BibitemShut {NoStop}%
\bibitem [{\citenamefont {Pan}\ \emph {et~al.}(2020{\natexlab{a}})\citenamefont
  {Pan}, \citenamefont {Wu},\ and\ \citenamefont {Sarma}}]{pan2020band}%
  \BibitemOpen
  \bibfield  {author} {\bibinfo {author} {\bibfnamefont {Haining}\ \bibnamefont
  {Pan}}, \bibinfo {author} {\bibfnamefont {Fengcheng}\ \bibnamefont {Wu}}, \
  and\ \bibinfo {author} {\bibfnamefont {Sankar~Das}\ \bibnamefont {Sarma}},\
  }\bibfield  {title} {\enquote {\bibinfo {title} {{Band topology, Hubbard
  model, Heisenberg model, and Dzyaloshinskii-Moriya interaction in twisted
  bilayer WSe$_2$}},}\ }\href@noop {} {\bibfield  {journal} {\bibinfo
  {journal} {Physical Review Research}\ }\textbf {\bibinfo {volume} {2}},\
  \bibinfo {pages} {033087} (\bibinfo {year} {2020}{\natexlab{a}})}\BibitemShut
  {NoStop}%
\bibitem [{\citenamefont {Pan}\ \emph {et~al.}(2020{\natexlab{b}})\citenamefont
  {Pan}, \citenamefont {Wu},\ and\ \citenamefont {Sarma}}]{pan2020quantum}%
  \BibitemOpen
  \bibfield  {author} {\bibinfo {author} {\bibfnamefont {Haining}\ \bibnamefont
  {Pan}}, \bibinfo {author} {\bibfnamefont {Fengcheng}\ \bibnamefont {Wu}}, \
  and\ \bibinfo {author} {\bibfnamefont {Sankar~Das}\ \bibnamefont {Sarma}},\
  }\bibfield  {title} {\enquote {\bibinfo {title} {{Quantum phase diagram of a
  Moir{\'e}-Hubbard model}},}\ }\href@noop {} {\bibfield  {journal} {\bibinfo
  {journal} {Physical Review B}\ }\textbf {\bibinfo {volume} {102}},\ \bibinfo
  {pages} {201104} (\bibinfo {year} {2020}{\natexlab{b}})}\BibitemShut
  {NoStop}%
\bibitem [{\citenamefont {Wietek}\ \emph {et~al.}(2022)\citenamefont {Wietek},
  \citenamefont {Wang}, \citenamefont {Zang}, \citenamefont {Cano},
  \citenamefont {Georges},\ and\ \citenamefont {Millis}}]{wietek2022tunable}%
  \BibitemOpen
  \bibfield  {author} {\bibinfo {author} {\bibfnamefont {Alexander}\
  \bibnamefont {Wietek}}, \bibinfo {author} {\bibfnamefont {Jie}\ \bibnamefont
  {Wang}}, \bibinfo {author} {\bibfnamefont {Jiawei}\ \bibnamefont {Zang}},
  \bibinfo {author} {\bibfnamefont {Jennifer}\ \bibnamefont {Cano}}, \bibinfo
  {author} {\bibfnamefont {Antoine}\ \bibnamefont {Georges}}, \ and\ \bibinfo
  {author} {\bibfnamefont {Andrew}\ \bibnamefont {Millis}},\ }\bibfield
  {title} {\enquote {\bibinfo {title} {{Tunable stripe order and weak
  superconductivity in the Moir{\'e} Hubbard model}},}\ }\href@noop {}
  {\bibfield  {journal} {\bibinfo  {journal} {Physical Review Research}\
  }\textbf {\bibinfo {volume} {4}},\ \bibinfo {pages} {043048} (\bibinfo {year}
  {2022})}\BibitemShut {NoStop}%
\bibitem [{\citenamefont {Morales-Dur{\'a}n}\ \emph {et~al.}(2022)\citenamefont
  {Morales-Dur{\'a}n}, \citenamefont {Hu}, \citenamefont {Potasz},\ and\
  \citenamefont {MacDonald}}]{morales2022nonlocal}%
  \BibitemOpen
  \bibfield  {author} {\bibinfo {author} {\bibfnamefont {Nicol{\'a}s}\
  \bibnamefont {Morales-Dur{\'a}n}}, \bibinfo {author} {\bibfnamefont
  {Nai~Chao}\ \bibnamefont {Hu}}, \bibinfo {author} {\bibfnamefont {Pawel}\
  \bibnamefont {Potasz}}, \ and\ \bibinfo {author} {\bibfnamefont {Allan~H}\
  \bibnamefont {MacDonald}},\ }\bibfield  {title} {\enquote {\bibinfo {title}
  {{Nonlocal interactions in moir{\'e} Hubbard systems}},}\ }\href@noop {}
  {\bibfield  {journal} {\bibinfo  {journal} {Physical Review Letters}\
  }\textbf {\bibinfo {volume} {128}},\ \bibinfo {pages} {217202} (\bibinfo
  {year} {2022})}\BibitemShut {NoStop}%
\bibitem [{\citenamefont {Huang}\ \emph {et~al.}(2019)\citenamefont {Huang},
  \citenamefont {Zhang},\ and\ \citenamefont
  {Ma}}]{huang2019antiferromagnetically}%
  \BibitemOpen
  \bibfield  {author} {\bibinfo {author} {\bibfnamefont {Tongyun}\ \bibnamefont
  {Huang}}, \bibinfo {author} {\bibfnamefont {Lufeng}\ \bibnamefont {Zhang}}, \
  and\ \bibinfo {author} {\bibfnamefont {Tianxing}\ \bibnamefont {Ma}},\
  }\bibfield  {title} {\enquote {\bibinfo {title} {{Antiferromagnetically
  ordered mott insulator and $d+id$ superconductivity in twisted bilayer
  graphene: A quantum monte carlo study}},}\ }\href@noop {} {\bibfield
  {journal} {\bibinfo  {journal} {Science Bulletin}\ }\textbf {\bibinfo
  {volume} {64}},\ \bibinfo {pages} {310--314} (\bibinfo {year}
  {2019})}\BibitemShut {NoStop}%
\bibitem [{\citenamefont {Lu}\ and\ \citenamefont
  {S{\'e}n{\'e}chal}(2022)}]{lu2022doping}%
  \BibitemOpen
  \bibfield  {author} {\bibinfo {author} {\bibfnamefont {Xiancong}\
  \bibnamefont {Lu}}\ and\ \bibinfo {author} {\bibfnamefont {David}\
  \bibnamefont {S{\'e}n{\'e}chal}},\ }\bibfield  {title} {\enquote {\bibinfo
  {title} {{Doping phase diagram of a Hubbard model for twisted bilayer
  cuprates}},}\ }\href@noop {} {\bibfield  {journal} {\bibinfo  {journal}
  {Physical Review B}\ }\textbf {\bibinfo {volume} {105}},\ \bibinfo {pages}
  {245127} (\bibinfo {year} {2022})}\BibitemShut {NoStop}%
\bibitem [{\citenamefont {B{\'e}langer}\ and\ \citenamefont
  {S{\'e}n{\'e}chal}(2023)}]{belanger2023doping}%
  \BibitemOpen
  \bibfield  {author} {\bibinfo {author} {\bibfnamefont {Mathieu}\ \bibnamefont
  {B{\'e}langer}}\ and\ \bibinfo {author} {\bibfnamefont {David}\ \bibnamefont
  {S{\'e}n{\'e}chal}},\ }\bibfield  {title} {\enquote {\bibinfo {title}
  {{Doping dependence of chiral superconductivity in near $45^\circ$ twisted
  bilayer cuprates}},}\ }\href@noop {} {\bibfield  {journal} {\bibinfo
  {journal} {arXiv preprint arXiv:2306.05485}\ } (\bibinfo {year}
  {2023})}\BibitemShut {NoStop}%
\bibitem [{\citenamefont {Fournier}\ \emph {et~al.}(2010)\citenamefont
  {Fournier}, \citenamefont {Levy}, \citenamefont {Pennec}, \citenamefont
  {McChesney}, \citenamefont {Bostwick}, \citenamefont {Rotenberg},
  \citenamefont {Liang}, \citenamefont {Hardy}, \citenamefont {Bonn},
  \citenamefont {Elfimov} \emph {et~al.}}]{fournier2010loss}%
  \BibitemOpen
  \bibfield  {author} {\bibinfo {author} {\bibfnamefont {D}~\bibnamefont
  {Fournier}}, \bibinfo {author} {\bibfnamefont {G}~\bibnamefont {Levy}},
  \bibinfo {author} {\bibfnamefont {Y}~\bibnamefont {Pennec}}, \bibinfo
  {author} {\bibfnamefont {JL}~\bibnamefont {McChesney}}, \bibinfo {author}
  {\bibfnamefont {A}~\bibnamefont {Bostwick}}, \bibinfo {author} {\bibfnamefont
  {E}~\bibnamefont {Rotenberg}}, \bibinfo {author} {\bibfnamefont
  {R}~\bibnamefont {Liang}}, \bibinfo {author} {\bibfnamefont {WN}~\bibnamefont
  {Hardy}}, \bibinfo {author} {\bibfnamefont {DA}~\bibnamefont {Bonn}},
  \bibinfo {author} {\bibfnamefont {IS}~\bibnamefont {Elfimov}},  \emph
  {et~al.},\ }\bibfield  {title} {\enquote {\bibinfo {title} {{Loss of nodal
  quasiparticle integrity in underdoped
  YBa$\rm{_2}$Cu$\rm{_3}$O$\rm{_{6+x}}$}},}\ }\href@noop {} {\bibfield
  {journal} {\bibinfo  {journal} {Nature Physics}\ }\textbf {\bibinfo {volume}
  {6}},\ \bibinfo {pages} {905--911} (\bibinfo {year} {2010})}\BibitemShut
  {NoStop}%
\bibitem [{\citenamefont {Gall}\ \emph {et~al.}(2021)\citenamefont {Gall},
  \citenamefont {Wurz}, \citenamefont {Samland}, \citenamefont {Chan},\ and\
  \citenamefont {K{\"o}hl}}]{gall2021competing}%
  \BibitemOpen
  \bibfield  {author} {\bibinfo {author} {\bibfnamefont {Marcell}\ \bibnamefont
  {Gall}}, \bibinfo {author} {\bibfnamefont {Nicola}\ \bibnamefont {Wurz}},
  \bibinfo {author} {\bibfnamefont {Jens}\ \bibnamefont {Samland}}, \bibinfo
  {author} {\bibfnamefont {Chun~Fai}\ \bibnamefont {Chan}}, \ and\ \bibinfo
  {author} {\bibfnamefont {Michael}\ \bibnamefont {K{\"o}hl}},\ }\bibfield
  {title} {\enquote {\bibinfo {title} {{Competing magnetic orders in a bilayer
  Hubbard model with ultracold atoms}},}\ }\href@noop {} {\bibfield  {journal}
  {\bibinfo  {journal} {Nature}\ }\textbf {\bibinfo {volume} {589}},\ \bibinfo
  {pages} {40--43} (\bibinfo {year} {2021})}\BibitemShut {NoStop}%
\bibitem [{\citenamefont {Scalettar}\ \emph {et~al.}(1994)\citenamefont
  {Scalettar}, \citenamefont {Cannon}, \citenamefont {Scalapino},\ and\
  \citenamefont {Sugar}}]{scalettar1994magnetic}%
  \BibitemOpen
  \bibfield  {author} {\bibinfo {author} {\bibfnamefont {Richard~T}\
  \bibnamefont {Scalettar}}, \bibinfo {author} {\bibfnamefont {Joel~W}\
  \bibnamefont {Cannon}}, \bibinfo {author} {\bibfnamefont {Douglas~J}\
  \bibnamefont {Scalapino}}, \ and\ \bibinfo {author} {\bibfnamefont
  {Robert~L}\ \bibnamefont {Sugar}},\ }\bibfield  {title} {\enquote {\bibinfo
  {title} {{Magnetic and pairing correlations in coupled Hubbard planes}},}\
  }\href@noop {} {\bibfield  {journal} {\bibinfo  {journal} {Physical Review
  B}\ }\textbf {\bibinfo {volume} {50}},\ \bibinfo {pages} {13419} (\bibinfo
  {year} {1994})}\BibitemShut {NoStop}%
\bibitem [{\citenamefont {Fuhrmann}\ \emph {et~al.}(2006)\citenamefont
  {Fuhrmann}, \citenamefont {Heilmann},\ and\ \citenamefont
  {Monien}}]{fuhrmann2006mott}%
  \BibitemOpen
  \bibfield  {author} {\bibinfo {author} {\bibfnamefont {Andreas}\ \bibnamefont
  {Fuhrmann}}, \bibinfo {author} {\bibfnamefont {David}\ \bibnamefont
  {Heilmann}}, \ and\ \bibinfo {author} {\bibfnamefont {Hartmut}\ \bibnamefont
  {Monien}},\ }\bibfield  {title} {\enquote {\bibinfo {title} {{From Mott
  insulator to band insulator: A dynamical mean-field theory study}},}\
  }\href@noop {} {\bibfield  {journal} {\bibinfo  {journal} {Physical Review
  B}\ }\textbf {\bibinfo {volume} {73}},\ \bibinfo {pages} {245118} (\bibinfo
  {year} {2006})}\BibitemShut {NoStop}%
\bibitem [{\citenamefont {Kancharla}\ and\ \citenamefont
  {Okamoto}(2007)}]{kancharla2007band}%
  \BibitemOpen
  \bibfield  {author} {\bibinfo {author} {\bibfnamefont
  {Srivenkateswara~Sarma}\ \bibnamefont {Kancharla}}\ and\ \bibinfo {author}
  {\bibfnamefont {Satoshi}\ \bibnamefont {Okamoto}},\ }\bibfield  {title}
  {\enquote {\bibinfo {title} {{Band insulator to Mott insulator transition in
  a bilayer Hubbard model}},}\ }\href@noop {} {\bibfield  {journal} {\bibinfo
  {journal} {Physical Review B}\ }\textbf {\bibinfo {volume} {75}},\ \bibinfo
  {pages} {193103} (\bibinfo {year} {2007})}\BibitemShut {NoStop}%
\bibitem [{\citenamefont {Bouadim}\ \emph {et~al.}(2008)\citenamefont
  {Bouadim}, \citenamefont {Batrouni}, \citenamefont {H{\'e}bert},\ and\
  \citenamefont {Scalettar}}]{bouadim2008magnetic}%
  \BibitemOpen
  \bibfield  {author} {\bibinfo {author} {\bibfnamefont {Karem}\ \bibnamefont
  {Bouadim}}, \bibinfo {author} {\bibfnamefont {G~George}\ \bibnamefont
  {Batrouni}}, \bibinfo {author} {\bibfnamefont {Frederic}\ \bibnamefont
  {H{\'e}bert}}, \ and\ \bibinfo {author} {\bibfnamefont {RT}~\bibnamefont
  {Scalettar}},\ }\bibfield  {title} {\enquote {\bibinfo {title} {{Magnetic and
  transport properties of a coupled Hubbard bilayer with electron and hole
  doping}},}\ }\href@noop {} {\bibfield  {journal} {\bibinfo  {journal}
  {Physical Review B}\ }\textbf {\bibinfo {volume} {77}},\ \bibinfo {pages}
  {144527} (\bibinfo {year} {2008})}\BibitemShut {NoStop}%
\bibitem [{\citenamefont {Mou}\ \emph {et~al.}(2022)\citenamefont {Mou},
  \citenamefont {Mondaini},\ and\ \citenamefont {Scalettar}}]{mou2022bilayer}%
  \BibitemOpen
  \bibfield  {author} {\bibinfo {author} {\bibfnamefont {Yingping}\
  \bibnamefont {Mou}}, \bibinfo {author} {\bibfnamefont {Rubem}\ \bibnamefont
  {Mondaini}}, \ and\ \bibinfo {author} {\bibfnamefont {Richard~T}\
  \bibnamefont {Scalettar}},\ }\bibfield  {title} {\enquote {\bibinfo {title}
  {{Bilayer Hubbard model: Analysis based on the fermionic sign problem}},}\
  }\href@noop {} {\bibfield  {journal} {\bibinfo  {journal} {Physical Review
  B}\ }\textbf {\bibinfo {volume} {106}},\ \bibinfo {pages} {125116} (\bibinfo
  {year} {2022})}\BibitemShut {NoStop}%
\bibitem [{\citenamefont {Golor}\ \emph {et~al.}(2014)\citenamefont {Golor},
  \citenamefont {Reckling}, \citenamefont {Classen}, \citenamefont {Scherer},\
  and\ \citenamefont {Wessel}}]{golor2014ground}%
  \BibitemOpen
  \bibfield  {author} {\bibinfo {author} {\bibfnamefont {Michael}\ \bibnamefont
  {Golor}}, \bibinfo {author} {\bibfnamefont {Timo}\ \bibnamefont {Reckling}},
  \bibinfo {author} {\bibfnamefont {Laura}\ \bibnamefont {Classen}}, \bibinfo
  {author} {\bibfnamefont {Michael~M}\ \bibnamefont {Scherer}}, \ and\ \bibinfo
  {author} {\bibfnamefont {Stefan}\ \bibnamefont {Wessel}},\ }\bibfield
  {title} {\enquote {\bibinfo {title} {{Ground-state phase diagram of the
  half-filled bilayer Hubbard model}},}\ }\href@noop {} {\bibfield  {journal}
  {\bibinfo  {journal} {Physical Review B}\ }\textbf {\bibinfo {volume} {90}},\
  \bibinfo {pages} {195131} (\bibinfo {year} {2014})}\BibitemShut {NoStop}%
\bibitem [{\citenamefont {R{\"u}ger}\ \emph {et~al.}(2014)\citenamefont
  {R{\"u}ger}, \citenamefont {Tocchio}, \citenamefont {Valent{\'\i}},\ and\
  \citenamefont {Gros}}]{ruger2014phase}%
  \BibitemOpen
  \bibfield  {author} {\bibinfo {author} {\bibfnamefont {Robert}\ \bibnamefont
  {R{\"u}ger}}, \bibinfo {author} {\bibfnamefont {Luca~F}\ \bibnamefont
  {Tocchio}}, \bibinfo {author} {\bibfnamefont {Roser}\ \bibnamefont
  {Valent{\'\i}}}, \ and\ \bibinfo {author} {\bibfnamefont {Claudius}\
  \bibnamefont {Gros}},\ }\bibfield  {title} {\enquote {\bibinfo {title} {{The
  phase diagram of the square lattice bilayer Hubbard model: a variational
  Monte Carlo study}},}\ }\href@noop {} {\bibfield  {journal} {\bibinfo
  {journal} {New Journal of Physics}\ }\textbf {\bibinfo {volume} {16}},\
  \bibinfo {pages} {033010} (\bibinfo {year} {2014})}\BibitemShut {NoStop}%
\bibitem [{\citenamefont {Lanat{\`a}}\ \emph {et~al.}(2009)\citenamefont
  {Lanat{\`a}}, \citenamefont {Barone},\ and\ \citenamefont
  {Fabrizio}}]{lanata2009superconductivity}%
  \BibitemOpen
  \bibfield  {author} {\bibinfo {author} {\bibfnamefont {Nicola}\ \bibnamefont
  {Lanat{\`a}}}, \bibinfo {author} {\bibfnamefont {Paolo}\ \bibnamefont
  {Barone}}, \ and\ \bibinfo {author} {\bibfnamefont {Michele}\ \bibnamefont
  {Fabrizio}},\ }\bibfield  {title} {\enquote {\bibinfo {title}
  {{Superconductivity in the doped bilayer Hubbard model}},}\ }\href@noop {}
  {\bibfield  {journal} {\bibinfo  {journal} {Physical Review B}\ }\textbf
  {\bibinfo {volume} {80}},\ \bibinfo {pages} {224524} (\bibinfo {year}
  {2009})}\BibitemShut {NoStop}%
\bibitem [{\citenamefont {Zhai}\ \emph {et~al.}(2009)\citenamefont {Zhai},
  \citenamefont {Wang},\ and\ \citenamefont
  {Lee}}]{zhai2009antiferromagnetically}%
  \BibitemOpen
  \bibfield  {author} {\bibinfo {author} {\bibfnamefont {Hui}\ \bibnamefont
  {Zhai}}, \bibinfo {author} {\bibfnamefont {Fa}~\bibnamefont {Wang}}, \ and\
  \bibinfo {author} {\bibfnamefont {Dung-Hai}\ \bibnamefont {Lee}},\ }\bibfield
   {title} {\enquote {\bibinfo {title} {{Antiferromagnetically driven
  electronic correlations in iron pnictides and cuprates}},}\ }\href@noop {}
  {\bibfield  {journal} {\bibinfo  {journal} {Physical Review B}\ }\textbf
  {\bibinfo {volume} {80}},\ \bibinfo {pages} {064517} (\bibinfo {year}
  {2009})}\BibitemShut {NoStop}%
\bibitem [{\citenamefont {Maier}\ and\ \citenamefont
  {Scalapino}(2011)}]{maier2011pair}%
  \BibitemOpen
  \bibfield  {author} {\bibinfo {author} {\bibfnamefont {Thomas~A}\
  \bibnamefont {Maier}}\ and\ \bibinfo {author} {\bibfnamefont
  {DJ}~\bibnamefont {Scalapino}},\ }\bibfield  {title} {\enquote {\bibinfo
  {title} {{Pair structure and the pairing interaction in a bilayer Hubbard
  model for unconventional superconductivity}},}\ }\href@noop {} {\bibfield
  {journal} {\bibinfo  {journal} {Physical Review B}\ }\textbf {\bibinfo
  {volume} {84}},\ \bibinfo {pages} {180513} (\bibinfo {year}
  {2011})}\BibitemShut {NoStop}%
\bibitem [{\citenamefont {Matsumoto}\ \emph {et~al.}(2020)\citenamefont
  {Matsumoto}, \citenamefont {Ogura},\ and\ \citenamefont
  {Kuroki}}]{matsumoto2020strongly}%
  \BibitemOpen
  \bibfield  {author} {\bibinfo {author} {\bibfnamefont {Karin}\ \bibnamefont
  {Matsumoto}}, \bibinfo {author} {\bibfnamefont {Daisuke}\ \bibnamefont
  {Ogura}}, \ and\ \bibinfo {author} {\bibfnamefont {Kazuhiko}\ \bibnamefont
  {Kuroki}},\ }\bibfield  {title} {\enquote {\bibinfo {title} {{Strongly
  enhanced superconductivity due to finite energy spin fluctuations induced by
  an incipient band: a flex study on the bilayer Hubbard model with vertical
  and diagonal interlayer hoppings}},}\ }\href@noop {} {\bibfield  {journal}
  {\bibinfo  {journal} {Journal of the Physical Society of Japan}\ }\textbf
  {\bibinfo {volume} {89}},\ \bibinfo {pages} {044709} (\bibinfo {year}
  {2020})}\BibitemShut {NoStop}%
\bibitem [{\citenamefont {Kato}\ and\ \citenamefont
  {Kuroki}(2020)}]{kato2020many}%
  \BibitemOpen
  \bibfield  {author} {\bibinfo {author} {\bibfnamefont {Daichi}\ \bibnamefont
  {Kato}}\ and\ \bibinfo {author} {\bibfnamefont {Kazuhiko}\ \bibnamefont
  {Kuroki}},\ }\bibfield  {title} {\enquote {\bibinfo {title} {{Many-variable
  variational Monte Carlo study of superconductivity in two-band Hubbard models
  with an incipient band}},}\ }\href@noop {} {\bibfield  {journal} {\bibinfo
  {journal} {Physical Review Research}\ }\textbf {\bibinfo {volume} {2}},\
  \bibinfo {pages} {023156} (\bibinfo {year} {2020})}\BibitemShut {NoStop}%
\bibitem [{\citenamefont {Karakuzu}\ \emph {et~al.}(2021)\citenamefont
  {Karakuzu}, \citenamefont {Johnston},\ and\ \citenamefont
  {Maier}}]{karakuzu2021superconductivity}%
  \BibitemOpen
  \bibfield  {author} {\bibinfo {author} {\bibfnamefont {Seher}\ \bibnamefont
  {Karakuzu}}, \bibinfo {author} {\bibfnamefont {Steven}\ \bibnamefont
  {Johnston}}, \ and\ \bibinfo {author} {\bibfnamefont {Thomas~A}\ \bibnamefont
  {Maier}},\ }\bibfield  {title} {\enquote {\bibinfo {title}
  {{Superconductivity in the bilayer Hubbard model: Two Fermi surfaces are
  better than one}},}\ }\href@noop {} {\bibfield  {journal} {\bibinfo
  {journal} {Physical Review B}\ }\textbf {\bibinfo {volume} {104}},\ \bibinfo
  {pages} {245109} (\bibinfo {year} {2021})}\BibitemShut {NoStop}%
\bibitem [{\citenamefont {Iwano}\ and\ \citenamefont
  {Yamaji}(2022)}]{iwano2022superconductivity}%
  \BibitemOpen
  \bibfield  {author} {\bibinfo {author} {\bibfnamefont {Akito}\ \bibnamefont
  {Iwano}}\ and\ \bibinfo {author} {\bibfnamefont {Youhei}\ \bibnamefont
  {Yamaji}},\ }\bibfield  {title} {\enquote {\bibinfo {title}
  {{Superconductivity in Bilayer t--t$^{\prime}$ Hubbard Models}},}\
  }\href@noop {} {\bibfield  {journal} {\bibinfo  {journal} {Journal of the
  Physical Society of Japan}\ }\textbf {\bibinfo {volume} {91}},\ \bibinfo
  {pages} {094702} (\bibinfo {year} {2022})}\BibitemShut {NoStop}%
\bibitem [{\citenamefont {Lee}\ \emph {et~al.}(2014)\citenamefont {Lee},
  \citenamefont {Zhang}, \citenamefont {Jeschke},\ and\ \citenamefont
  {Valent{\'\i}}}]{lee2014competition}%
  \BibitemOpen
  \bibfield  {author} {\bibinfo {author} {\bibfnamefont {Hunpyo}\ \bibnamefont
  {Lee}}, \bibinfo {author} {\bibfnamefont {Yu-Zhong}\ \bibnamefont {Zhang}},
  \bibinfo {author} {\bibfnamefont {Harald~O}\ \bibnamefont {Jeschke}}, \ and\
  \bibinfo {author} {\bibfnamefont {Roser}\ \bibnamefont {Valent{\'\i}}},\
  }\bibfield  {title} {\enquote {\bibinfo {title} {{Competition between band
  and Mott insulators in the bilayer Hubbard model: A dynamical cluster
  approximation study}},}\ }\href@noop {} {\bibfield  {journal} {\bibinfo
  {journal} {Physical Review B}\ }\textbf {\bibinfo {volume} {89}},\ \bibinfo
  {pages} {035139} (\bibinfo {year} {2014})}\BibitemShut {NoStop}%
\bibitem [{\citenamefont {Vanhala}\ \emph {et~al.}(2015)\citenamefont
  {Vanhala}, \citenamefont {Baarsma}, \citenamefont {Heikkinen}, \citenamefont
  {Troyer}, \citenamefont {Harju},\ and\ \citenamefont
  {T{\"o}rm{\"a}}}]{vanhala2015superfluidity}%
  \BibitemOpen
  \bibfield  {author} {\bibinfo {author} {\bibfnamefont {Tuomas~I}\
  \bibnamefont {Vanhala}}, \bibinfo {author} {\bibfnamefont {Jildou~E}\
  \bibnamefont {Baarsma}}, \bibinfo {author} {\bibfnamefont {Miikka~OJ}\
  \bibnamefont {Heikkinen}}, \bibinfo {author} {\bibfnamefont {Matthias}\
  \bibnamefont {Troyer}}, \bibinfo {author} {\bibfnamefont {Ari}\ \bibnamefont
  {Harju}}, \ and\ \bibinfo {author} {\bibfnamefont {P{\"a}ivi}\ \bibnamefont
  {T{\"o}rm{\"a}}},\ }\bibfield  {title} {\enquote {\bibinfo {title}
  {{Superfluidity and density order in a bilayer extended Hubbard model}},}\
  }\href@noop {} {\bibfield  {journal} {\bibinfo  {journal} {Physical Review
  B}\ }\textbf {\bibinfo {volume} {91}},\ \bibinfo {pages} {144510} (\bibinfo
  {year} {2015})}\BibitemShut {NoStop}%
\bibitem [{\citenamefont {Park}\ \emph {et~al.}(2012)\citenamefont {Park},
  \citenamefont {Millis},\ and\ \citenamefont {Marianetti}}]{park2012site}%
  \BibitemOpen
  \bibfield  {author} {\bibinfo {author} {\bibfnamefont {Hyowon}\ \bibnamefont
  {Park}}, \bibinfo {author} {\bibfnamefont {Andrew~J}\ \bibnamefont {Millis}},
  \ and\ \bibinfo {author} {\bibfnamefont {Chris~A}\ \bibnamefont
  {Marianetti}},\ }\bibfield  {title} {\enquote {\bibinfo {title}
  {{Site-selective Mott transition in rare-earth-element nickelates}},}\
  }\href@noop {} {\bibfield  {journal} {\bibinfo  {journal} {Physical review
  letters}\ }\textbf {\bibinfo {volume} {109}},\ \bibinfo {pages} {156402}
  (\bibinfo {year} {2012})}\BibitemShut {NoStop}%
\bibitem [{\citenamefont {Shimizu}\ \emph {et~al.}(2015)\citenamefont
  {Shimizu}, \citenamefont {Aoyama}, \citenamefont {Jinno}, \citenamefont
  {Itoh},\ and\ \citenamefont {Ueda}}]{shimizu2015site}%
  \BibitemOpen
  \bibfield  {author} {\bibinfo {author} {\bibfnamefont {Yasuhiro}\
  \bibnamefont {Shimizu}}, \bibinfo {author} {\bibfnamefont {Satoshi}\
  \bibnamefont {Aoyama}}, \bibinfo {author} {\bibfnamefont {Takaaki}\
  \bibnamefont {Jinno}}, \bibinfo {author} {\bibfnamefont {Masayuki}\
  \bibnamefont {Itoh}}, \ and\ \bibinfo {author} {\bibfnamefont {Yutaka}\
  \bibnamefont {Ueda}},\ }\bibfield  {title} {\enquote {\bibinfo {title}
  {{Site-Selective Mott Transition in a Quasi-One-Dimensional Vanadate
  V$\rm{_6}$O$\rm{_{13}}$}},}\ }\href@noop {} {\bibfield  {journal} {\bibinfo
  {journal} {Physical Review Letters}\ }\textbf {\bibinfo {volume} {114}},\
  \bibinfo {pages} {166403} (\bibinfo {year} {2015})}\BibitemShut {NoStop}%
\bibitem [{\citenamefont {Yu}\ \emph {et~al.}(2014)\citenamefont {Yu},
  \citenamefont {Liu}, \citenamefont {Quan}, \citenamefont {Jia}, \citenamefont
  {Lin},\ and\ \citenamefont {Zou}}]{yu2014site}%
  \BibitemOpen
  \bibfield  {author} {\bibinfo {author} {\bibfnamefont {Xiang-Long}\
  \bibnamefont {Yu}}, \bibinfo {author} {\bibfnamefont {Da-Yong}\ \bibnamefont
  {Liu}}, \bibinfo {author} {\bibfnamefont {Ya-Min}\ \bibnamefont {Quan}},
  \bibinfo {author} {\bibfnamefont {Ting}\ \bibnamefont {Jia}}, \bibinfo
  {author} {\bibfnamefont {Hai-Qing}\ \bibnamefont {Lin}}, \ and\ \bibinfo
  {author} {\bibfnamefont {Liang-Jian}\ \bibnamefont {Zou}},\ }\bibfield
  {title} {\enquote {\bibinfo {title} {{A site-selective antiferromagnetic
  ground state in layered pnictide-oxide BaTi$\rm{_2}$As$\rm{_2}$O}},}\
  }\href@noop {} {\bibfield  {journal} {\bibinfo  {journal} {Journal of Applied
  Physics}\ }\textbf {\bibinfo {volume} {115}} (\bibinfo {year}
  {2014})}\BibitemShut {NoStop}%
\bibitem [{\citenamefont {Hubbard}(1963)}]{Hubbard1963Electron1}%
  \BibitemOpen
  \bibfield  {author} {\bibinfo {author} {\bibfnamefont {John}\ \bibnamefont
  {Hubbard}},\ }\bibfield  {title} {\enquote {\bibinfo {title} {Electron
  correlations in narrow energy bands},}\ }\href {\doibase
  10.1098/rspa.1963.0204} {\bibfield  {journal} {\bibinfo  {journal}
  {Proceedings of the Royal Society of London. Series A. Mathematical and
  Physical Sciences}\ }\textbf {\bibinfo {volume} {276}},\ \bibinfo {pages}
  {238--257} (\bibinfo {year} {1963})}\BibitemShut {NoStop}%
\bibitem [{\citenamefont {Soven}(1967)}]{soven1967coherent}%
  \BibitemOpen
  \bibfield  {author} {\bibinfo {author} {\bibfnamefont {Paul}\ \bibnamefont
  {Soven}},\ }\bibfield  {title} {\enquote {\bibinfo {title}
  {{Coherent-potential model of substitutional disordered alloys}},}\
  }\href@noop {} {\bibfield  {journal} {\bibinfo  {journal} {Physical Review}\
  }\textbf {\bibinfo {volume} {156}},\ \bibinfo {pages} {809} (\bibinfo {year}
  {1967})}\BibitemShut {NoStop}%
\bibitem [{\citenamefont {Velick{\`y}}\ \emph {et~al.}(1968)\citenamefont
  {Velick{\`y}}, \citenamefont {Kirkpatrick},\ and\ \citenamefont
  {Ehrenreich}}]{velicky1968single}%
  \BibitemOpen
  \bibfield  {author} {\bibinfo {author} {\bibfnamefont {B}~\bibnamefont
  {Velick{\`y}}}, \bibinfo {author} {\bibfnamefont {S}~\bibnamefont
  {Kirkpatrick}}, \ and\ \bibinfo {author} {\bibfnamefont {H}~\bibnamefont
  {Ehrenreich}},\ }\bibfield  {title} {\enquote {\bibinfo {title} {{Single-site
  approximations in the electronic theory of simple binary alloys}},}\
  }\href@noop {} {\bibfield  {journal} {\bibinfo  {journal} {Physical Review}\
  }\textbf {\bibinfo {volume} {175}},\ \bibinfo {pages} {747} (\bibinfo {year}
  {1968})}\BibitemShut {NoStop}%
\bibitem [{\citenamefont {Elliott}\ \emph {et~al.}(1974)\citenamefont
  {Elliott}, \citenamefont {Krumhansl},\ and\ \citenamefont
  {Leath}}]{elliott1974theory}%
  \BibitemOpen
  \bibfield  {author} {\bibinfo {author} {\bibfnamefont {RJ}~\bibnamefont
  {Elliott}}, \bibinfo {author} {\bibfnamefont {JA}~\bibnamefont {Krumhansl}},
  \ and\ \bibinfo {author} {\bibfnamefont {PL}~\bibnamefont {Leath}},\
  }\bibfield  {title} {\enquote {\bibinfo {title} {{The theory and properties
  of randomly disordered crystals and related physical systems}},}\ }\href@noop
  {} {\bibfield  {journal} {\bibinfo  {journal} {Reviews of modern physics}\
  }\textbf {\bibinfo {volume} {46}},\ \bibinfo {pages} {465} (\bibinfo {year}
  {1974})}\BibitemShut {NoStop}%
\bibitem [{\citenamefont {Gebhard}\ and\ \citenamefont
  {Gebhard}(1997)}]{gebhard1997metal}%
  \BibitemOpen
  \bibfield  {author} {\bibinfo {author} {\bibfnamefont {Florian}\ \bibnamefont
  {Gebhard}}\ and\ \bibinfo {author} {\bibfnamefont {Florian}\ \bibnamefont
  {Gebhard}},\ }\href@noop {} {\emph {\bibinfo {title} {{Metal—insulator
  transitions}}}}\ (\bibinfo  {publisher} {Springer},\ \bibinfo {year}
  {1997})\BibitemShut {NoStop}%
\bibitem [{\citenamefont {Hoang}(2010)}]{hoang2010metal}%
  \BibitemOpen
  \bibfield  {author} {\bibinfo {author} {\bibfnamefont {AT}~\bibnamefont
  {Hoang}},\ }\bibfield  {title} {\enquote {\bibinfo {title} {{Metal--insulator
  transitions in the half-filled ionic Hubbard model}},}\ }\href@noop {}
  {\bibfield  {journal} {\bibinfo  {journal} {Journal of physics: Condensed
  matter}\ }\textbf {\bibinfo {volume} {22}},\ \bibinfo {pages} {095602}
  (\bibinfo {year} {2010})}\BibitemShut {NoStop}%
\bibitem [{\citenamefont {Rowlands}\ and\ \citenamefont
  {Zhang}(2014{\natexlab{a}})}]{rowlands2014inclusion}%
  \BibitemOpen
  \bibfield  {author} {\bibinfo {author} {\bibfnamefont {DA}~\bibnamefont
  {Rowlands}}\ and\ \bibinfo {author} {\bibfnamefont {Yu-Zhong}\ \bibnamefont
  {Zhang}},\ }\bibfield  {title} {\enquote {\bibinfo {title} {{Inclusion of
  intersite spatial correlations in the alloy analogy approach to the
  half-filled ionic Hubbard model}},}\ }\href@noop {} {\bibfield  {journal}
  {\bibinfo  {journal} {Journal of Physics: Condensed Matter}\ }\textbf
  {\bibinfo {volume} {26}},\ \bibinfo {pages} {274201} (\bibinfo {year}
  {2014}{\natexlab{a}})}\BibitemShut {NoStop}%
\bibitem [{\citenamefont {Rowlands}\ and\ \citenamefont
  {Zhang}(2014{\natexlab{b}})}]{rowlands2014disappearance}%
  \BibitemOpen
  \bibfield  {author} {\bibinfo {author} {\bibfnamefont {DA}~\bibnamefont
  {Rowlands}}\ and\ \bibinfo {author} {\bibfnamefont {Yu-Zhong}\ \bibnamefont
  {Zhang}},\ }\bibfield  {title} {\enquote {\bibinfo {title} {{Disappearance of
  the Dirac cone in silicene due to the presence of an electric field}},}\
  }\href@noop {} {\bibfield  {journal} {\bibinfo  {journal} {Chinese Physics
  B}\ }\textbf {\bibinfo {volume} {23}},\ \bibinfo {pages} {037101} (\bibinfo
  {year} {2014}{\natexlab{b}})}\BibitemShut {NoStop}%
\bibitem [{\citenamefont {Assaad}\ and\ \citenamefont
  {Herbut}(2013)}]{assaad2013pinning}%
  \BibitemOpen
  \bibfield  {author} {\bibinfo {author} {\bibfnamefont {Fakher~F}\
  \bibnamefont {Assaad}}\ and\ \bibinfo {author} {\bibfnamefont {Igor~F}\
  \bibnamefont {Herbut}},\ }\bibfield  {title} {\enquote {\bibinfo {title}
  {{Pinning the order: the nature of quantum criticality in the Hubbard model
  on honeycomb lattice}},}\ }\href@noop {} {\bibfield  {journal} {\bibinfo
  {journal} {Physical Review X}\ }\textbf {\bibinfo {volume} {3}},\ \bibinfo
  {pages} {031010} (\bibinfo {year} {2013})}\BibitemShut {NoStop}%
\bibitem [{\citenamefont {Sorella}\ \emph {et~al.}(2012)\citenamefont
  {Sorella}, \citenamefont {Otsuka},\ and\ \citenamefont
  {Yunoki}}]{sorella2012absence}%
  \BibitemOpen
  \bibfield  {author} {\bibinfo {author} {\bibfnamefont {Sandro}\ \bibnamefont
  {Sorella}}, \bibinfo {author} {\bibfnamefont {Yuichi}\ \bibnamefont
  {Otsuka}}, \ and\ \bibinfo {author} {\bibfnamefont {Seiji}\ \bibnamefont
  {Yunoki}},\ }\bibfield  {title} {\enquote {\bibinfo {title} {{Absence of a
  spin liquid phase in the Hubbard model on the honeycomb lattice}},}\
  }\href@noop {} {\bibfield  {journal} {\bibinfo  {journal} {Scientific
  reports}\ }\textbf {\bibinfo {volume} {2}},\ \bibinfo {pages} {992} (\bibinfo
  {year} {2012})}\BibitemShut {NoStop}%
\bibitem [{\citenamefont {Toldin}\ \emph {et~al.}(2015)\citenamefont {Toldin},
  \citenamefont {Hohenadler}, \citenamefont {Assaad},\ and\ \citenamefont
  {Herbut}}]{toldin2015fermionic}%
  \BibitemOpen
  \bibfield  {author} {\bibinfo {author} {\bibfnamefont {Francesco~Parisen}\
  \bibnamefont {Toldin}}, \bibinfo {author} {\bibfnamefont {Martin}\
  \bibnamefont {Hohenadler}}, \bibinfo {author} {\bibfnamefont {Fakher~F}\
  \bibnamefont {Assaad}}, \ and\ \bibinfo {author} {\bibfnamefont {Igor~F}\
  \bibnamefont {Herbut}},\ }\bibfield  {title} {\enquote {\bibinfo {title}
  {{Fermionic quantum criticality in honeycomb and $\pi$-flux Hubbard models:
  Finite-size scaling of renormalization-group-invariant observables from
  quantum Monte Carlo}},}\ }\href@noop {} {\bibfield  {journal} {\bibinfo
  {journal} {Physical Review B}\ }\textbf {\bibinfo {volume} {91}},\ \bibinfo
  {pages} {165108} (\bibinfo {year} {2015})}\BibitemShut {NoStop}%
\bibitem [{\citenamefont {Wu}\ \emph {et~al.}(2010)\citenamefont {Wu},
  \citenamefont {Chen}, \citenamefont {Tao}, \citenamefont {Tong},\ and\
  \citenamefont {Liu}}]{wu2010interacting}%
  \BibitemOpen
  \bibfield  {author} {\bibinfo {author} {\bibfnamefont {Wei}\ \bibnamefont
  {Wu}}, \bibinfo {author} {\bibfnamefont {Yao-Hua}\ \bibnamefont {Chen}},
  \bibinfo {author} {\bibfnamefont {Hong-Shuai}\ \bibnamefont {Tao}}, \bibinfo
  {author} {\bibfnamefont {Ning-Hua}\ \bibnamefont {Tong}}, \ and\ \bibinfo
  {author} {\bibfnamefont {Wu-Ming}\ \bibnamefont {Liu}},\ }\bibfield  {title}
  {\enquote {\bibinfo {title} {{Interacting Dirac fermions on honeycomb
  lattice}},}\ }\href@noop {} {\bibfield  {journal} {\bibinfo  {journal}
  {Physical Review B}\ }\textbf {\bibinfo {volume} {82}},\ \bibinfo {pages}
  {245102} (\bibinfo {year} {2010})}\BibitemShut {NoStop}%
\bibitem [{\citenamefont {Liebsch}(2011)}]{liebsch2011correlated}%
  \BibitemOpen
  \bibfield  {author} {\bibinfo {author} {\bibfnamefont {Ansgar}\ \bibnamefont
  {Liebsch}},\ }\bibfield  {title} {\enquote {\bibinfo {title} {{Correlated
  Dirac fermions on the honeycomb lattice studied within cluster dynamical mean
  field theory}},}\ }\href@noop {} {\bibfield  {journal} {\bibinfo  {journal}
  {Physical Review B}\ }\textbf {\bibinfo {volume} {83}},\ \bibinfo {pages}
  {035113} (\bibinfo {year} {2011})}\BibitemShut {NoStop}%
\bibitem [{\citenamefont {Mak}\ \emph {et~al.}(2009)\citenamefont {Mak},
  \citenamefont {Lui}, \citenamefont {Shan},\ and\ \citenamefont
  {Heinz}}]{mak2009observation}%
  \BibitemOpen
  \bibfield  {author} {\bibinfo {author} {\bibfnamefont {Kin~Fai}\ \bibnamefont
  {Mak}}, \bibinfo {author} {\bibfnamefont {Chun~Hung}\ \bibnamefont {Lui}},
  \bibinfo {author} {\bibfnamefont {Jie}\ \bibnamefont {Shan}}, \ and\ \bibinfo
  {author} {\bibfnamefont {Tony~F}\ \bibnamefont {Heinz}},\ }\bibfield  {title}
  {\enquote {\bibinfo {title} {{Observation of an electric-field-induced band
  gap in bilayer graphene by infrared spectroscopy}},}\ }\href@noop {}
  {\bibfield  {journal} {\bibinfo  {journal} {Physical review letters}\
  }\textbf {\bibinfo {volume} {102}},\ \bibinfo {pages} {256405} (\bibinfo
  {year} {2009})}\BibitemShut {NoStop}%
\bibitem [{\citenamefont {Weitz}\ \emph {et~al.}(2010)\citenamefont {Weitz},
  \citenamefont {Allen}, \citenamefont {Feldman}, \citenamefont {Martin},\ and\
  \citenamefont {Yacoby}}]{weitz2010broken}%
  \BibitemOpen
  \bibfield  {author} {\bibinfo {author} {\bibfnamefont {R~Thomas}\
  \bibnamefont {Weitz}}, \bibinfo {author} {\bibfnamefont {Monica~T}\
  \bibnamefont {Allen}}, \bibinfo {author} {\bibfnamefont {Benjamin~E}\
  \bibnamefont {Feldman}}, \bibinfo {author} {\bibfnamefont {Jens}\
  \bibnamefont {Martin}}, \ and\ \bibinfo {author} {\bibfnamefont {Amir}\
  \bibnamefont {Yacoby}},\ }\bibfield  {title} {\enquote {\bibinfo {title}
  {{Broken-symmetry states in doubly gated suspended bilayer graphene}},}\
  }\href@noop {} {\bibfield  {journal} {\bibinfo  {journal} {Science}\ }\textbf
  {\bibinfo {volume} {330}},\ \bibinfo {pages} {812--816} (\bibinfo {year}
  {2010})}\BibitemShut {NoStop}%
\bibitem [{\citenamefont {Hunt}\ \emph {et~al.}(2013)\citenamefont {Hunt},
  \citenamefont {Sanchez-Yamagishi}, \citenamefont {Young}, \citenamefont
  {Yankowitz}, \citenamefont {LeRoy}, \citenamefont {Watanabe}, \citenamefont
  {Taniguchi}, \citenamefont {Moon}, \citenamefont {Koshino}, \citenamefont
  {Jarillo-Herrero} \emph {et~al.}}]{hunt2013massive}%
  \BibitemOpen
  \bibfield  {author} {\bibinfo {author} {\bibfnamefont {Benjamin}\
  \bibnamefont {Hunt}}, \bibinfo {author} {\bibfnamefont {Javier~D}\
  \bibnamefont {Sanchez-Yamagishi}}, \bibinfo {author} {\bibfnamefont
  {Andrea~F}\ \bibnamefont {Young}}, \bibinfo {author} {\bibfnamefont
  {Matthew}\ \bibnamefont {Yankowitz}}, \bibinfo {author} {\bibfnamefont
  {Brian~J}\ \bibnamefont {LeRoy}}, \bibinfo {author} {\bibfnamefont {Kenji}\
  \bibnamefont {Watanabe}}, \bibinfo {author} {\bibfnamefont {Takashi}\
  \bibnamefont {Taniguchi}}, \bibinfo {author} {\bibfnamefont {Pilkyung}\
  \bibnamefont {Moon}}, \bibinfo {author} {\bibfnamefont {Mikito}\ \bibnamefont
  {Koshino}}, \bibinfo {author} {\bibfnamefont {Pablo}\ \bibnamefont
  {Jarillo-Herrero}},  \emph {et~al.},\ }\bibfield  {title} {\enquote {\bibinfo
  {title} {{Massive Dirac fermions and Hofstadter butterfly in a van der Waals
  heterostructure}},}\ }\href@noop {} {\bibfield  {journal} {\bibinfo
  {journal} {Science}\ }\textbf {\bibinfo {volume} {340}},\ \bibinfo {pages}
  {1427--1430} (\bibinfo {year} {2013})}\BibitemShut {NoStop}%
\bibitem [{\citenamefont {Chen}\ \emph {et~al.}(2014)\citenamefont {Chen},
  \citenamefont {Shi}, \citenamefont {Yang}, \citenamefont {Lu}, \citenamefont
  {Lai}, \citenamefont {Yan}, \citenamefont {Wang}, \citenamefont {Zhang},\
  and\ \citenamefont {Li}}]{chen2014observation}%
  \BibitemOpen
  \bibfield  {author} {\bibinfo {author} {\bibfnamefont {Zhi-Guo}\ \bibnamefont
  {Chen}}, \bibinfo {author} {\bibfnamefont {Zhiwen}\ \bibnamefont {Shi}},
  \bibinfo {author} {\bibfnamefont {Wei}\ \bibnamefont {Yang}}, \bibinfo
  {author} {\bibfnamefont {Xiaobo}\ \bibnamefont {Lu}}, \bibinfo {author}
  {\bibfnamefont {You}\ \bibnamefont {Lai}}, \bibinfo {author} {\bibfnamefont
  {Hugen}\ \bibnamefont {Yan}}, \bibinfo {author} {\bibfnamefont {Feng}\
  \bibnamefont {Wang}}, \bibinfo {author} {\bibfnamefont {Guangyu}\
  \bibnamefont {Zhang}}, \ and\ \bibinfo {author} {\bibfnamefont {Zhiqiang}\
  \bibnamefont {Li}},\ }\bibfield  {title} {\enquote {\bibinfo {title}
  {{Observation of an intrinsic bandgap and Landau level renormalization in
  graphene/boron-nitride heterostructures}},}\ }\href@noop {} {\bibfield
  {journal} {\bibinfo  {journal} {Nature communications}\ }\textbf {\bibinfo
  {volume} {5}},\ \bibinfo {pages} {4461} (\bibinfo {year} {2014})}\BibitemShut
  {NoStop}%
\bibitem [{\citenamefont {Xu}\ \emph {et~al.}(2016{\natexlab{a}})\citenamefont
  {Xu}, \citenamefont {Song}, \citenamefont {Lin},\ and\ \citenamefont
  {Zhang}}]{xu2016gate}%
  \BibitemOpen
  \bibfield  {author} {\bibinfo {author} {\bibfnamefont {Jin-Rong}\
  \bibnamefont {Xu}}, \bibinfo {author} {\bibfnamefont {Ze-Yi}\ \bibnamefont
  {Song}}, \bibinfo {author} {\bibfnamefont {Hai-Qing}\ \bibnamefont {Lin}}, \
  and\ \bibinfo {author} {\bibfnamefont {Yu-Zhong}\ \bibnamefont {Zhang}},\
  }\bibfield  {title} {\enquote {\bibinfo {title} {{Gate-induced gap in bilayer
  graphene suppressed by Coulomb repulsion}},}\ }\href@noop {} {\bibfield
  {journal} {\bibinfo  {journal} {Physical Review B}\ }\textbf {\bibinfo
  {volume} {93}},\ \bibinfo {pages} {035109} (\bibinfo {year}
  {2016}{\natexlab{a}})}\BibitemShut {NoStop}%
\bibitem [{\citenamefont {Xu}\ \emph {et~al.}(2016{\natexlab{b}})\citenamefont
  {Xu}, \citenamefont {Song}, \citenamefont {Yuan},\ and\ \citenamefont
  {Zhang}}]{xu2016interaction}%
  \BibitemOpen
  \bibfield  {author} {\bibinfo {author} {\bibfnamefont {Jin-Rong}\
  \bibnamefont {Xu}}, \bibinfo {author} {\bibfnamefont {Ze-Yi}\ \bibnamefont
  {Song}}, \bibinfo {author} {\bibfnamefont {Chen-Guang}\ \bibnamefont {Yuan}},
  \ and\ \bibinfo {author} {\bibfnamefont {Yu-Zhong}\ \bibnamefont {Zhang}},\
  }\bibfield  {title} {\enquote {\bibinfo {title} {{Interaction-induced
  metallic state in graphene on hexagonal boron nitride}},}\ }\href@noop {}
  {\bibfield  {journal} {\bibinfo  {journal} {Physical Review B}\ }\textbf
  {\bibinfo {volume} {94}},\ \bibinfo {pages} {195103} (\bibinfo {year}
  {2016}{\natexlab{b}})}\BibitemShut {NoStop}%
\bibitem [{\citenamefont {Groeber}\ \emph {et~al.}(1998)\citenamefont
  {Groeber}, \citenamefont {Zacher},\ and\ \citenamefont
  {Eder}}]{groeber1998paramagnetic}%
  \BibitemOpen
  \bibfield  {author} {\bibinfo {author} {\bibfnamefont {C}~\bibnamefont
  {Groeber}}, \bibinfo {author} {\bibfnamefont {MG}~\bibnamefont {Zacher}}, \
  and\ \bibinfo {author} {\bibfnamefont {R}~\bibnamefont {Eder}},\ }\bibfield
  {title} {\enquote {\bibinfo {title} {{Paramagnetic metal-insulator transition
  in the 2D Hubbard model}},}\ }\href@noop {} {\bibfield  {journal} {\bibinfo
  {journal} {arXiv preprint cond-mat/9810246}\ } (\bibinfo {year}
  {1998})}\BibitemShut {NoStop}%
\bibitem [{\citenamefont {Gull}\ \emph {et~al.}(2013)\citenamefont {Gull},
  \citenamefont {Parcollet},\ and\ \citenamefont
  {Millis}}]{gull2013superconductivity}%
  \BibitemOpen
  \bibfield  {author} {\bibinfo {author} {\bibfnamefont {Emanuel}\ \bibnamefont
  {Gull}}, \bibinfo {author} {\bibfnamefont {Olivier}\ \bibnamefont
  {Parcollet}}, \ and\ \bibinfo {author} {\bibfnamefont {Andrew~J}\
  \bibnamefont {Millis}},\ }\bibfield  {title} {\enquote {\bibinfo {title}
  {{Superconductivity and the pseudogap in the two-dimensional Hubbard
  model}},}\ }\href@noop {} {\bibfield  {journal} {\bibinfo  {journal}
  {Physical review letters}\ }\textbf {\bibinfo {volume} {110}},\ \bibinfo
  {pages} {216405} (\bibinfo {year} {2013})}\BibitemShut {NoStop}%
\bibitem [{\citenamefont {Song}\ \emph {et~al.}(2017)\citenamefont {Song},
  \citenamefont {Jiang}, \citenamefont {Lin},\ and\ \citenamefont
  {Zhang}}]{song2017distinct}%
  \BibitemOpen
  \bibfield  {author} {\bibinfo {author} {\bibfnamefont {Ze-Yi}\ \bibnamefont
  {Song}}, \bibinfo {author} {\bibfnamefont {Xiu-Cai}\ \bibnamefont {Jiang}},
  \bibinfo {author} {\bibfnamefont {Hai-Qing}\ \bibnamefont {Lin}}, \ and\
  \bibinfo {author} {\bibfnamefont {Yu-Zhong}\ \bibnamefont {Zhang}},\
  }\bibfield  {title} {\enquote {\bibinfo {title} {{Distinct nature of
  orbital-selective Mott phases dominated by low-energy local spin
  fluctuations}},}\ }\href@noop {} {\bibfield  {journal} {\bibinfo  {journal}
  {Physical Review B}\ }\textbf {\bibinfo {volume} {96}},\ \bibinfo {pages}
  {235119} (\bibinfo {year} {2017})}\BibitemShut {NoStop}%
\bibitem [{\citenamefont {Kakehashi}\ and\ \citenamefont
  {Fulde}(2004)}]{kakehashi2004coherent}%
  \BibitemOpen
  \bibfield  {author} {\bibinfo {author} {\bibfnamefont {Y}~\bibnamefont
  {Kakehashi}}\ and\ \bibinfo {author} {\bibfnamefont {P}~\bibnamefont
  {Fulde}},\ }\bibfield  {title} {\enquote {\bibinfo {title} {{Coherent
  potential approximation and projection operators for interacting
  electrons}},}\ }\href@noop {} {\bibfield  {journal} {\bibinfo  {journal}
  {Physical Review B}\ }\textbf {\bibinfo {volume} {69}},\ \bibinfo {pages}
  {045101} (\bibinfo {year} {2004})}\BibitemShut {NoStop}%
\bibitem [{\citenamefont {Garg}\ \emph {et~al.}(2006)\citenamefont {Garg},
  \citenamefont {Krishnamurthy},\ and\ \citenamefont {Randeria}}]{garg2006can}%
  \BibitemOpen
  \bibfield  {author} {\bibinfo {author} {\bibfnamefont {Arti}\ \bibnamefont
  {Garg}}, \bibinfo {author} {\bibfnamefont {HR}~\bibnamefont {Krishnamurthy}},
  \ and\ \bibinfo {author} {\bibfnamefont {Mohit}\ \bibnamefont {Randeria}},\
  }\bibfield  {title} {\enquote {\bibinfo {title} {{Can correlations drive a
  band insulator metallic?}}}\ }\href@noop {} {\bibfield  {journal} {\bibinfo
  {journal} {Physical review letters}\ }\textbf {\bibinfo {volume} {97}},\
  \bibinfo {pages} {046403} (\bibinfo {year} {2006})}\BibitemShut {NoStop}%
\bibitem [{\citenamefont {Luo}\ and\ \citenamefont
  {Wang}(2000)}]{luo2000higher}%
  \BibitemOpen
  \bibfield  {author} {\bibinfo {author} {\bibfnamefont {Hong-Gang}\
  \bibnamefont {Luo}}\ and\ \bibinfo {author} {\bibfnamefont {Shun-Jin}\
  \bibnamefont {Wang}},\ }\bibfield  {title} {\enquote {\bibinfo {title}
  {{Higher-order correlation effects to the solution of the Hubbard model}},}\
  }\href@noop {} {\bibfield  {journal} {\bibinfo  {journal} {Physical Review
  B}\ }\textbf {\bibinfo {volume} {61}},\ \bibinfo {pages} {5158} (\bibinfo
  {year} {2000})}\BibitemShut {NoStop}%
\bibitem [{\citenamefont {Kotliar}\ \emph {et~al.}(2001)\citenamefont
  {Kotliar}, \citenamefont {Savrasov}, \citenamefont {P{\'a}lsson},\ and\
  \citenamefont {Biroli}}]{kotliar2001cellular}%
  \BibitemOpen
  \bibfield  {author} {\bibinfo {author} {\bibfnamefont {Gabriel}\ \bibnamefont
  {Kotliar}}, \bibinfo {author} {\bibfnamefont {Sergej~Y}\ \bibnamefont
  {Savrasov}}, \bibinfo {author} {\bibfnamefont {Gunnar}\ \bibnamefont
  {P{\'a}lsson}}, \ and\ \bibinfo {author} {\bibfnamefont {Giulio}\
  \bibnamefont {Biroli}},\ }\bibfield  {title} {\enquote {\bibinfo {title}
  {{Cellular dynamical mean field approach to strongly correlated systems}},}\
  }\href@noop {} {\bibfield  {journal} {\bibinfo  {journal} {Physical review
  letters}\ }\textbf {\bibinfo {volume} {87}},\ \bibinfo {pages} {186401}
  (\bibinfo {year} {2001})}\BibitemShut {NoStop}%
\bibitem [{\citenamefont {Xu}\ \emph {et~al.}(2022)\citenamefont {Xu},
  \citenamefont {Ray}, \citenamefont {Shao}, \citenamefont {Jiang},
  \citenamefont {Lee}, \citenamefont {Weber}, \citenamefont {Goldberger},
  \citenamefont {Watanabe}, \citenamefont {Taniguchi}, \citenamefont {Muller}
  \emph {et~al.}}]{xu2022coexisting}%
  \BibitemOpen
  \bibfield  {author} {\bibinfo {author} {\bibfnamefont {Yang}\ \bibnamefont
  {Xu}}, \bibinfo {author} {\bibfnamefont {Ariana}\ \bibnamefont {Ray}},
  \bibinfo {author} {\bibfnamefont {Yu-Tsun}\ \bibnamefont {Shao}}, \bibinfo
  {author} {\bibfnamefont {Shengwei}\ \bibnamefont {Jiang}}, \bibinfo {author}
  {\bibfnamefont {Kihong}\ \bibnamefont {Lee}}, \bibinfo {author}
  {\bibfnamefont {Daniel}\ \bibnamefont {Weber}}, \bibinfo {author}
  {\bibfnamefont {Joshua~E}\ \bibnamefont {Goldberger}}, \bibinfo {author}
  {\bibfnamefont {Kenji}\ \bibnamefont {Watanabe}}, \bibinfo {author}
  {\bibfnamefont {Takashi}\ \bibnamefont {Taniguchi}}, \bibinfo {author}
  {\bibfnamefont {David~A}\ \bibnamefont {Muller}},  \emph {et~al.},\
  }\bibfield  {title} {\enquote {\bibinfo {title} {{Coexisting
  ferromagnetic--antiferromagnetic state in twisted bilayer CrI$_3$}},}\
  }\href@noop {} {\bibfield  {journal} {\bibinfo  {journal} {Nature
  Nanotechnology}\ }\textbf {\bibinfo {volume} {17}},\ \bibinfo {pages}
  {143--147} (\bibinfo {year} {2022})}\BibitemShut {NoStop}%
\bibitem [{\citenamefont {Pant}\ and\ \citenamefont
  {Pati}(2022)}]{pant2022phase}%
  \BibitemOpen
  \bibfield  {author} {\bibinfo {author} {\bibfnamefont {Dharmendra}\
  \bibnamefont {Pant}}\ and\ \bibinfo {author} {\bibfnamefont {Ranjit}\
  \bibnamefont {Pati}},\ }\bibfield  {title} {\enquote {\bibinfo {title}
  {{Phase transition from a nonmagnetic to a ferromagnetic state in a twisted
  bilayer graphene nanoflake: The role of electronic pressure on the
  magic-twist}},}\ }\href@noop {} {\bibfield  {journal} {\bibinfo  {journal}
  {Nanoscale}\ }\textbf {\bibinfo {volume} {14}},\ \bibinfo {pages}
  {11945--11952} (\bibinfo {year} {2022})}\BibitemShut {NoStop}%
\bibitem [{\citenamefont {Yu}\ \emph {et~al.}(2019)\citenamefont {Yu},
  \citenamefont {Ma}, \citenamefont {Cai}, \citenamefont {Zhong}, \citenamefont
  {Ye}, \citenamefont {Shen}, \citenamefont {Gu}, \citenamefont {Chen},\ and\
  \citenamefont {Zhang}}]{yu2019high}%
  \BibitemOpen
  \bibfield  {author} {\bibinfo {author} {\bibfnamefont {Yijun}\ \bibnamefont
  {Yu}}, \bibinfo {author} {\bibfnamefont {Liguo}\ \bibnamefont {Ma}}, \bibinfo
  {author} {\bibfnamefont {Peng}\ \bibnamefont {Cai}}, \bibinfo {author}
  {\bibfnamefont {Ruidan}\ \bibnamefont {Zhong}}, \bibinfo {author}
  {\bibfnamefont {Cun}\ \bibnamefont {Ye}}, \bibinfo {author} {\bibfnamefont
  {Jian}\ \bibnamefont {Shen}}, \bibinfo {author} {\bibfnamefont {Genda~D}\
  \bibnamefont {Gu}}, \bibinfo {author} {\bibfnamefont {Xian~Hui}\ \bibnamefont
  {Chen}}, \ and\ \bibinfo {author} {\bibfnamefont {Yuanbo}\ \bibnamefont
  {Zhang}},\ }\bibfield  {title} {\enquote {\bibinfo {title} {{High-temperature
  superconductivity in monolayer
  Bi$\rm{_2}$Sr$\rm{_2}$CaCu$\rm{_2}$O$\rm{_{8+\delta}}$}},}\ }\href@noop {}
  {\bibfield  {journal} {\bibinfo  {journal} {Nature}\ }\textbf {\bibinfo
  {volume} {575}},\ \bibinfo {pages} {156--163} (\bibinfo {year}
  {2019})}\BibitemShut {NoStop}%
\bibitem [{\citenamefont {Zhao}\ \emph {et~al.}(2019)\citenamefont {Zhao},
  \citenamefont {Poccia}, \citenamefont {Panetta}, \citenamefont {Yu},
  \citenamefont {Johnson}, \citenamefont {Yoo}, \citenamefont {Zhong},
  \citenamefont {Gu}, \citenamefont {Watanabe}, \citenamefont {Taniguchi} \emph
  {et~al.}}]{zhao2019sign}%
  \BibitemOpen
  \bibfield  {author} {\bibinfo {author} {\bibfnamefont {SY~Frank}\
  \bibnamefont {Zhao}}, \bibinfo {author} {\bibfnamefont {Nicola}\ \bibnamefont
  {Poccia}}, \bibinfo {author} {\bibfnamefont {Margaret~G}\ \bibnamefont
  {Panetta}}, \bibinfo {author} {\bibfnamefont {Cyndia}\ \bibnamefont {Yu}},
  \bibinfo {author} {\bibfnamefont {Jedediah~W}\ \bibnamefont {Johnson}},
  \bibinfo {author} {\bibfnamefont {Hyobin}\ \bibnamefont {Yoo}}, \bibinfo
  {author} {\bibfnamefont {Ruidan}\ \bibnamefont {Zhong}}, \bibinfo {author}
  {\bibfnamefont {GD}~\bibnamefont {Gu}}, \bibinfo {author} {\bibfnamefont
  {Kenji}\ \bibnamefont {Watanabe}}, \bibinfo {author} {\bibfnamefont
  {Takashi}\ \bibnamefont {Taniguchi}},  \emph {et~al.},\ }\bibfield  {title}
  {\enquote {\bibinfo {title} {{Sign-reversing Hall effect in atomically thin
  high-temperature
  Bi$\rm{_{2.1}}$Sr$\rm{_{1.9}}$CaCu$\rm{_2}$O$\rm{_{8+\delta}}$
  superconductors}},}\ }\href@noop {} {\bibfield  {journal} {\bibinfo
  {journal} {Physical review letters}\ }\textbf {\bibinfo {volume} {122}},\
  \bibinfo {pages} {247001} (\bibinfo {year} {2019})}\BibitemShut {NoStop}%
\bibitem [{\citenamefont {Li}\ \emph {et~al.}(1999)\citenamefont {Li},
  \citenamefont {Tsay}, \citenamefont {Suenaga}, \citenamefont {Klemm},
  \citenamefont {Gu},\ and\ \citenamefont {Koshizuka}}]{li1999bi}%
  \BibitemOpen
  \bibfield  {author} {\bibinfo {author} {\bibfnamefont {Qiang}\ \bibnamefont
  {Li}}, \bibinfo {author} {\bibfnamefont {YN}~\bibnamefont {Tsay}}, \bibinfo
  {author} {\bibfnamefont {M}~\bibnamefont {Suenaga}}, \bibinfo {author}
  {\bibfnamefont {RA}~\bibnamefont {Klemm}}, \bibinfo {author} {\bibfnamefont
  {GD}~\bibnamefont {Gu}}, \ and\ \bibinfo {author} {\bibfnamefont
  {N}~\bibnamefont {Koshizuka}},\ }\bibfield  {title} {\enquote {\bibinfo
  {title} {{Bi$\rm{_2}$Sr$\rm{_2}$CaCu$\rm{_2}$O$\rm{_{8+\delta}}$ bicrystal
  c-axis twist Josephson junctions: a new phase-sensitive test of order
  parameter symmetry}},}\ }\href@noop {} {\bibfield  {journal} {\bibinfo
  {journal} {Physical review letters}\ }\textbf {\bibinfo {volume} {83}},\
  \bibinfo {pages} {4160} (\bibinfo {year} {1999})}\BibitemShut {NoStop}%
\bibitem [{\citenamefont {Takano}\ \emph {et~al.}(2002)\citenamefont {Takano},
  \citenamefont {Hatano}, \citenamefont {Fukuyo}, \citenamefont {Ishii},
  \citenamefont {Ohmori}, \citenamefont {Arisawa}, \citenamefont {Togano},\
  and\ \citenamefont {Tachiki}}]{takano2002d}%
  \BibitemOpen
  \bibfield  {author} {\bibinfo {author} {\bibfnamefont {Yoshihiko}\
  \bibnamefont {Takano}}, \bibinfo {author} {\bibfnamefont {Takeshi}\
  \bibnamefont {Hatano}}, \bibinfo {author} {\bibfnamefont {Akihiro}\
  \bibnamefont {Fukuyo}}, \bibinfo {author} {\bibfnamefont {Akira}\
  \bibnamefont {Ishii}}, \bibinfo {author} {\bibfnamefont {Masashi}\
  \bibnamefont {Ohmori}}, \bibinfo {author} {\bibfnamefont {Shunichi}\
  \bibnamefont {Arisawa}}, \bibinfo {author} {\bibfnamefont {Kazumasa}\
  \bibnamefont {Togano}}, \ and\ \bibinfo {author} {\bibfnamefont {Masashi}\
  \bibnamefont {Tachiki}},\ }\bibfield  {title} {\enquote {\bibinfo {title}
  {{d-like symmetry of the order parameter and intrinsic Josephson effects in
  Bi$\rm{_2}$Sr$\rm{_2}$CaCu$\rm{_2}$O$\rm{_{8+\delta}}$ cross-whisker
  junctions}},}\ }\href@noop {} {\bibfield  {journal} {\bibinfo  {journal}
  {Physical Review B}\ }\textbf {\bibinfo {volume} {65}},\ \bibinfo {pages}
  {140513} (\bibinfo {year} {2002})}\BibitemShut {NoStop}%
\bibitem [{\citenamefont {Latyshev}\ \emph {et~al.}(2004)\citenamefont
  {Latyshev}, \citenamefont {Orlov}, \citenamefont {Nikitina}, \citenamefont
  {Monceau},\ and\ \citenamefont {Klemm}}]{latyshev2004c}%
  \BibitemOpen
  \bibfield  {author} {\bibinfo {author} {\bibfnamefont {Yu~I}\ \bibnamefont
  {Latyshev}}, \bibinfo {author} {\bibfnamefont {AP}~\bibnamefont {Orlov}},
  \bibinfo {author} {\bibfnamefont {AM}~\bibnamefont {Nikitina}}, \bibinfo
  {author} {\bibfnamefont {P}~\bibnamefont {Monceau}}, \ and\ \bibinfo {author}
  {\bibfnamefont {RA}~\bibnamefont {Klemm}},\ }\bibfield  {title} {\enquote
  {\bibinfo {title} {{c-axis transport in naturally grown
  Bi$\rm{_2}$Sr$\rm{_2}$CaCu$\rm{_2}$O$\rm{_{8+\delta}}$ cross-whisker
  junctions}},}\ }\href@noop {} {\bibfield  {journal} {\bibinfo  {journal}
  {Physical Review B}\ }\textbf {\bibinfo {volume} {70}},\ \bibinfo {pages}
  {094517} (\bibinfo {year} {2004})}\BibitemShut {NoStop}%
\bibitem [{\citenamefont {Yang}\ \emph {et~al.}(2018)\citenamefont {Yang},
  \citenamefont {Qin}, \citenamefont {Zhang}, \citenamefont {Fang},\ and\
  \citenamefont {Hu}}]{yang2018pi}%
  \BibitemOpen
  \bibfield  {author} {\bibinfo {author} {\bibfnamefont {Zhesen}\ \bibnamefont
  {Yang}}, \bibinfo {author} {\bibfnamefont {Shengshan}\ \bibnamefont {Qin}},
  \bibinfo {author} {\bibfnamefont {Qiang}\ \bibnamefont {Zhang}}, \bibinfo
  {author} {\bibfnamefont {Chen}\ \bibnamefont {Fang}}, \ and\ \bibinfo
  {author} {\bibfnamefont {Jiangping}\ \bibnamefont {Hu}},\ }\bibfield  {title}
  {\enquote {\bibinfo {title} {{$\pi$/2-Josephson junction as a topological
  superconductor}},}\ }\href@noop {} {\bibfield  {journal} {\bibinfo  {journal}
  {Physical Review B}\ }\textbf {\bibinfo {volume} {98}},\ \bibinfo {pages}
  {104515} (\bibinfo {year} {2018})}\BibitemShut {NoStop}%
\bibitem [{\citenamefont {Zhu}\ \emph {et~al.}(2021)\citenamefont {Zhu},
  \citenamefont {Liao}, \citenamefont {Zhang}, \citenamefont {Xie},
  \citenamefont {Meng}, \citenamefont {Liu}, \citenamefont {Bai}, \citenamefont
  {Ji}, \citenamefont {Zhang}, \citenamefont {Jiang} \emph
  {et~al.}}]{zhu2021presence}%
  \BibitemOpen
  \bibfield  {author} {\bibinfo {author} {\bibfnamefont {Yuying}\ \bibnamefont
  {Zhu}}, \bibinfo {author} {\bibfnamefont {Menghan}\ \bibnamefont {Liao}},
  \bibinfo {author} {\bibfnamefont {Qinghua}\ \bibnamefont {Zhang}}, \bibinfo
  {author} {\bibfnamefont {Hong-Yi}\ \bibnamefont {Xie}}, \bibinfo {author}
  {\bibfnamefont {Fanqi}\ \bibnamefont {Meng}}, \bibinfo {author}
  {\bibfnamefont {Yaowu}\ \bibnamefont {Liu}}, \bibinfo {author} {\bibfnamefont
  {Zhonghua}\ \bibnamefont {Bai}}, \bibinfo {author} {\bibfnamefont {Shuaihua}\
  \bibnamefont {Ji}}, \bibinfo {author} {\bibfnamefont {Jin}\ \bibnamefont
  {Zhang}}, \bibinfo {author} {\bibfnamefont {Kaili}\ \bibnamefont {Jiang}},
  \emph {et~al.},\ }\bibfield  {title} {\enquote {\bibinfo {title} {{Presence
  of s-wave pairing in Josephson junctions made of twisted ultrathin
  Bi$\rm{_2}$Sr$\rm{_2}$CaCu$\rm{_2}$O$\rm{_{8+x}}$ flakes}},}\ }\href@noop {}
  {\bibfield  {journal} {\bibinfo  {journal} {Physical Review X}\ }\textbf
  {\bibinfo {volume} {11}},\ \bibinfo {pages} {031011} (\bibinfo {year}
  {2021})}\BibitemShut {NoStop}%
\bibitem [{\citenamefont {Zhao}\ \emph {et~al.}(2021)\citenamefont {Zhao},
  \citenamefont {Poccia}, \citenamefont {Cui}, \citenamefont {Volkov},
  \citenamefont {Yoo}, \citenamefont {Engelke}, \citenamefont {Ronen},
  \citenamefont {Zhong}, \citenamefont {Gu}, \citenamefont {Plugge} \emph
  {et~al.}}]{zhao2021emergent}%
  \BibitemOpen
  \bibfield  {author} {\bibinfo {author} {\bibfnamefont {SY}~\bibnamefont
  {Zhao}}, \bibinfo {author} {\bibfnamefont {Nicola}\ \bibnamefont {Poccia}},
  \bibinfo {author} {\bibfnamefont {Xiaomeng}\ \bibnamefont {Cui}}, \bibinfo
  {author} {\bibfnamefont {Pavel~A}\ \bibnamefont {Volkov}}, \bibinfo {author}
  {\bibfnamefont {Hyobin}\ \bibnamefont {Yoo}}, \bibinfo {author}
  {\bibfnamefont {Rebecca}\ \bibnamefont {Engelke}}, \bibinfo {author}
  {\bibfnamefont {Yuval}\ \bibnamefont {Ronen}}, \bibinfo {author}
  {\bibfnamefont {Ruidan}\ \bibnamefont {Zhong}}, \bibinfo {author}
  {\bibfnamefont {Genda}\ \bibnamefont {Gu}}, \bibinfo {author} {\bibfnamefont
  {Stephan}\ \bibnamefont {Plugge}},  \emph {et~al.},\ }\bibfield  {title}
  {\enquote {\bibinfo {title} {{Emergent Interfacial Superconductivity between
  Twisted Cuprate Superconductors}},}\ }\href@noop {} {\bibfield  {journal}
  {\bibinfo  {journal} {arXiv e-prints}\ ,\ \bibinfo {pages} {arXiv--2108}}
  (\bibinfo {year} {2021})}\BibitemShut {NoStop}%
\bibitem [{\citenamefont {Lee}\ \emph {et~al.}(2021)\citenamefont {Lee},
  \citenamefont {Lee}, \citenamefont {Kim}, \citenamefont {Choi}, \citenamefont
  {Park}, \citenamefont {Jang}, \citenamefont {Gu}, \citenamefont {Choi},
  \citenamefont {Cho}, \citenamefont {Lee} \emph {et~al.}}]{lee2021twisted}%
  \BibitemOpen
  \bibfield  {author} {\bibinfo {author} {\bibfnamefont {Jongyun}\ \bibnamefont
  {Lee}}, \bibinfo {author} {\bibfnamefont {Wonjun}\ \bibnamefont {Lee}},
  \bibinfo {author} {\bibfnamefont {Gi-Yeop}\ \bibnamefont {Kim}}, \bibinfo
  {author} {\bibfnamefont {Yong-Bin}\ \bibnamefont {Choi}}, \bibinfo {author}
  {\bibfnamefont {Jinho}\ \bibnamefont {Park}}, \bibinfo {author}
  {\bibfnamefont {Seong}\ \bibnamefont {Jang}}, \bibinfo {author}
  {\bibfnamefont {Genda}\ \bibnamefont {Gu}}, \bibinfo {author} {\bibfnamefont
  {Si-Young}\ \bibnamefont {Choi}}, \bibinfo {author} {\bibfnamefont
  {Gil~Young}\ \bibnamefont {Cho}}, \bibinfo {author} {\bibfnamefont {Gil-Ho}\
  \bibnamefont {Lee}},  \emph {et~al.},\ }\bibfield  {title} {\enquote
  {\bibinfo {title} {{Twisted van der Waals Josephson Junction Based on a
  High-Tc Superconductor}},}\ }\href@noop {} {\bibfield  {journal} {\bibinfo
  {journal} {Nano Letters}\ }\textbf {\bibinfo {volume} {21}},\ \bibinfo
  {pages} {10469--10477} (\bibinfo {year} {2021})}\BibitemShut {NoStop}%
\bibitem [{\citenamefont {Trambly~de Laissardi{\`e}re}\ \emph
  {et~al.}(2010)\citenamefont {Trambly~de Laissardi{\`e}re}, \citenamefont
  {Mayou},\ and\ \citenamefont {Magaud}}]{trambly2010localization}%
  \BibitemOpen
  \bibfield  {author} {\bibinfo {author} {\bibfnamefont {G}~\bibnamefont
  {Trambly~de Laissardi{\`e}re}}, \bibinfo {author} {\bibfnamefont {Didier}\
  \bibnamefont {Mayou}}, \ and\ \bibinfo {author} {\bibfnamefont {Laurence}\
  \bibnamefont {Magaud}},\ }\bibfield  {title} {\enquote {\bibinfo {title}
  {{Localization of Dirac electrons in rotated graphene bilayers}},}\
  }\href@noop {} {\bibfield  {journal} {\bibinfo  {journal} {Nano letters}\
  }\textbf {\bibinfo {volume} {10}},\ \bibinfo {pages} {804--808} (\bibinfo
  {year} {2010})}\BibitemShut {NoStop}%
\end{thebibliography}%

\end{document}